\documentclass[aps,prl]{revtex4}
\usepackage{amsmath}
\usepackage{amsfonts}
\usepackage{amssymb}
\usepackage{graphicx}
\usepackage{setspace}
\usepackage[english]{babel}

\newcommand\epigraph[3]{
\vspace{0.5em}\hfill{}\begin{minipage}{#1}{
\begin{spacing}{0.9}
\small\noindent\textit{#2}
\end{spacing}
\vspace{0.5em}
\hfill{}{#3}}\vspace{1em}
\end{minipage}}

\begin{document}

\title{Expanding Space, Quasars and St. Augustine's Fireworks}

\author{Olga I.~Chashchina}
\email{chashchina.olga@gmail.com}
\affiliation{\'{E}cole Polytechnique, 91128, Palaiseau, France}

\author{Zurab K.~Silagadze}
\email{Z.K.Silagadze@inp.nsk.su}
\affiliation{Budker Institute of Nuclear Physics and Novosibirsk State 
University, Novosibirsk 630 090, Russia }

\begin{abstract}
An attempt is made to explain time non-dilation allegedly observed
in quasar light curves. The explanation is based on the assumption that
quasar black holes are, in some sense, foreign for our
Friedmann-Robertson-Walker universe and do not participate in the Hubble
flow. Although at first sight such a weird explanation requires unreasonably
fine-tuned Big Bang initial conditions, we find a natural justification for
it using the Milne cosmological model as an inspiration.
\end{abstract}

\maketitle

\vspace{-10pt}

                              \epigraph{11cm}{\begin{verse}
                               You'd think capricious Hebe,\\
                               feeding the eagle of Zeus,\\
                               had raised a thunder-foaming goblet,\\
                               unable to restrain her mirth,\\
                               and tipped it on the earth.
                               \end{verse}}
                               {F.I.Tyutchev. A Spring Storm, 1828.   
                               Translated by F.Jude \cite{E}.}

\section{Introduction}

``Quasar light curves do not show the effects of time dilation''---this result
of the paper \cite{15} seems incredible. The cosmological time dilation is a
very basic phenomenon predicted by the Friedmann-Robertson-Walker (FRW)
metric. How can quasars ignore this fundamental effect of relativistic
cosmology?

Of course, extraordinary claims require extraordinary evidence. Until
the puzzling results of \cite{15} are firmly confirmed by independent
observations, we cannot be certain that what is observed in \cite{15} is
a real observation of time non-dilation in quasar light curves and not
a some unaccounted systematic error, for example, in data analysis.

Nevertheless, we will assume in this article that the results of \cite{15}
are correct and try to explain why quasars behave like strangers in our
FRW world. Strangely enough, our main idea came from Fyodor Tyutchev's
poetry (see the epigraph). The image of capricious Hebe tipping
a quasar-foaming goblet on our universe was a poetic metaphor
guiding this investigation and what follows can be considered as an attempt
of more or less scientific incarnation of this irresistible metaphor.

The paper is organized in five parts. In the first section we consider caveats
of the standard picture of expanding space as a rather subjective notion.
On the other hand, in contrast to expanding space, which can be considered as
a kind of coordinate artifact, cosmological time dilation is argued to be
an unavoidable, coordinate independent prediction for any internal time
variability of any object participating in the Hubble flow. Therefore, if
quasars really manage to escape bonds of time dilation, a natural inference
will be that they somehow do not participate in the Hubble flow. This quasar
mystery is accompanied with some others that are discussed in the second
section.

The following section gives a detailed description of the Milne model. In the
ideal case, the Milne universe is a part of the Minkowski space-time with the
future light-cone of the Big Bang event as its impenetrable boundary. However,
the density of the Milne fundamental observers, tending to infinity on this
boundary, makes unrealistic the realization of the ideal Milne universe. This
conclusion is further strengthened if we consider a quantum scalar field in
the Milne universe: if the boundary is assumed to be impenetrable, the
corresponding choice of the initial vacuum state is not an adiabatic but a
conformal vacuum that leads to the pair production phenomenon indicating the
presence of physically unrealistic infinite power sources on the Milne
universe's singular boundary.

In the forth section it is argued that in any realistic incarnation of the
Milne universe it is expected that objects from ``outside'' can penetrate
inside the Milne universe. Then we postulate that the existence of analogous
objects in our universe is possible and we comment how these objects which
do not participate in the Hubble flow due to their ``otherworldly'' origin,
can offer an explanation of some quasar mysteries.

In the final section we provide concluding remarks and discuss how the result
of this work can be related to other ideas described in the literature.

We have tried to make the paper as self-contained as possible by providing
enough details of all calculations, and we hope that ``the reader will not
need to have his fingers at eleven places to follow an~argument'' \cite{I-1}.

\section{Space Expansion and Enigma of Time Non-Dilation in Quasar Light
Curves}

Expansion of space is, probably, the most familiar concept in
modern cosmology. However, this beguilingly simple idea harbors many dangers
of misunderstanding and misuse even for professional physicists, ``most
scientists think they understand it, but few agree on what it really means''
\cite{1}. It is not surprising that the subtle notion of expanding space
became a subject of continuing debates \cite{1-A,1-B,1-C,1-D,1-E,1-F,1-G,1-H,
1-I,1-J,1-K}, largely triggered  by a beautiful {\it Scientific American}
article by Lineweaver and Davis \cite{1}.

We can identify at least two main reasons why it is dangerous and misleading
to speak about expanding space without explicitly clarifying what is really
meant by this combination of words.

First of all, as special relativity teaches us, space by itself does not
constitute an objective \mbox{reality---only} space-time does. As eloquently
expressed by Minkowski ``Henceforth, space by itself and time by itself are
doomed to fade away into mere shadows, and only a kind of union of the two
will preserve an independent reality'' \cite{2}. It does not make sense,
therefore, to speak about expanding space without clarifying what foliation
of space-time we have in mind and why we have chosen this \mbox{specific
foliation.}

In general relativity understanding is further complicated by the fact that
coordinates loose their usual meaning as temporal and spatial
labels of events. We can use any coordinate system in general
relativity to describe a given physical phenomenon and we should be very
careful to separate the real physical events from mere coordinate artifacts.
Moreover, when we speak about temporal development in space of some
physical process, we should specify what coordinate system was chosen and for
what reason. Without doing so, ``to some degree we mislead both our students
and ourselves when we calculate, for instance, the mercury perihelion motion
without explaining how our coordinate system is fixed in space, what
defines it in such a way that it cannot be rotated, by a few seconds a year,
to follow the perihelion's apparent motion \ldots Expressing our results in
terms of the values of coordinates became a habit to such an extend,
that we adhere to this habit also in general relativity, where values of
coordinates are not meaningful per se'' \cite{3}.

It is the symmetry of space-time, namely isotropy and homogeneity of space,
which determines its preferred foliation. More precisely, the cosmological
or Copernican  principle states that we do not occupy a special place in
the universe, which, at large, is very much the same everywhere at any given
instance. Mathematically this means (see, for example, \cite{4}) that the
space-time can be foliated into spacelike slices and has a form
$\mathbb{R}\times\Sigma$, where every spacelike slice $\Sigma$ is a maximally
symmetric space (homogeneous and isotropic) and $\mathbb{R}$
represents the cosmic time (a succession of instances of ``now'').
Geometrically the cosmic time is essentially the parameter that labels
spacelike slices and each slice hypersurface is an orbit of the symmetry
group of spatial homogeneity and isotropy \cite{5}.

The metric which corresponds to this preferred foliation is the celebrated
Friedmann-Robertson- Walker metric (in units where $c=1$):

\vspace{-12pt}
\begin{equation}
ds^2=dt^2-a^2(t)\left [\frac{dr^2}{1-kr^2}+r^2(d\theta^2+\sin^2{\theta}
d\phi^2)\right ],
\label{eq1}
\end{equation}
with $k>0$, $k=0$ or $k<0$ corresponding to the close, flat or open universes
respectively.

The expansion of space has a clear meaning in FRW coordinates: the scale
factor $a(t)$ depends on cosmic time. The particular form of this dependence
is determined from Einstein equations. However, this fact does not mean much
without connecting it to real physical observables: as was mentioned above,
we should distinguish real physical effects from coordinate artifacts.

One coordinate artifact is the singularity (for positive $k$) of Eq. 
(\ref{eq1}) at $r=1/\sqrt{k}$ and it can be removed by introducing a new 
radial coordinate $\psi$ according to \cite{6}:
\begin{equation}
r=S(\psi;k),
\label{eq2}
\end{equation}
where the generalized sine function $S(\psi;k)$ is defined as follows
\cite{7,8}:
\begin{equation}
S(\psi;k)=\left\{\begin{array}{c}
\frac{\sinh{(\sqrt{-k}\psi)}}{\sqrt{-k}},\;\, \mathrm{if} \;\,
k<0 \\ \psi, \hspace*{12mm}\;\; \mathrm{if} \hspace*{2mm} k=0 \\
\frac{\sin{(\sqrt{k}\psi)}}{\sqrt{k}},\;\; \mathrm{if} \;\;
k>0 \end{array} \right . .
\label{eq3}
\end{equation}
In new coordinates the metric takes the form \cite{6}
\begin{equation}
ds^2=dt^2-a^2(t)\left [d\psi^2+S^2(\psi;k)(d\theta^2+\sin^2{\theta}
d\phi^2)\right ].
\label{eq4}
\end{equation}

How can we be sure that the time dependence of the scale factor cannot be
similarly removed by a suitable change of coordinates? For example, let us
consider the following cosmological model. The metric Eq. (\ref{eq1}) 
has two characteristic length scales both of which in general evolve in time: 
the spatial curvature radius $R_c=a/\sqrt{6|k|}$ and the Hubble length
$R_H=H^{-1}$, where $H=\dot{a}/a$ is the Hubble parameter that sets the time
scale of the cosmic expansion. An attractive possibility is the perfect
cosmological principle that the universe at large is not only homogeneous
in space but also in time. Bondi's derisive formulation \cite{6}
``Geography does not matter, and history does not matter either'' expresses
the perfect cosmological principle more eloquently. It is immediately obvious
that for the non-dependence on time of the curvature radius and of the Hubble
length, it is necessary to have $k=0$ and $a(t)=a_0e^{Ht}$ with some constant
$a_0$ which can be absorbed into the exponential part and then eliminated by
a time translation (by  suitable redefinition of the origin of time scale).
Thus, the metric takes the form:
\begin{equation}
ds^2=dt^2-e^{2Ht}\left [dr^2+r^2(d\theta^2+\sin^2{\theta}d\phi^2)\right ].
\label{eq5}
\end{equation}
This beautiful theory, based on the perfect cosmological principle, used to
be a base for the steady state theory (see, for example, \cite{9}) before
the latter being killed by  some ugly observational facts \cite{10}. However,
nowadays this de Sitter cosmology surprisingly resurrected as a possible
description of reality as it may correspond to universe's initial
inflationary phase or its terminal state.

Let us introduce the new coordinates $T, R$ instead of $t, r$ through
relations (see, for example, \cite{1-F}):
\begin{equation}
t=T+\frac{1}{2H}\ln{\left (1-\frac{R^2}{R_H^2}\right )},\;\;\;
r=e^{-Ht}R=\frac{e^{-HT}}{\sqrt{1-\frac{R^2}{R_H^2}}}R.
\label{eq6}
\end{equation}

In this so called static coordinates the metric Eq. (\ref{eq5}) becomes
\begin{equation}
ds^2=\left (1-\frac{R^2}{R_H^2}\right )dT^2-\left (1-\frac{R^2}{R_H^2}
\right )^{-1}dR^2-R^2(d\theta^2+\sin^2{\theta}d\phi^2).
\label{eq7}
\end{equation}

The static coordinates determine another slicing (foliation) of the de Sitter
space, which doesn't show any sign of space expansion. What about other
foliations?

Let us introduce the following fancy coordinatization of the de Sitter space:
\begin{eqnarray} &&
x_0=R_H\left (\sinh{(Ht)}+\frac{r^2}{2R_H^2}e^{Ht} \right ),\;
x_4=R_H\left (\cosh{(Ht)}-\frac{r^2}{2R_H^2}e^{Ht} \right ),
\nonumber \\ &&
x_1=re^{Ht}\sin{\theta}\cos{\phi},\;\;\;\;\;
x_2=re^{Ht}\sin{\theta}\sin{\phi},\;\;\;\;\;
x_3=re^{Ht}\cos{\theta}.
\label{eq8}
\end{eqnarray}

This set of five coordinates is obviously not independent as the de Sitter
space-time is four dimensional. From Eq. (\ref{eq8}) we easily find that
\begin{equation}
-x_0^2+x_1^2+x_2^2+x_3^2+x_4^2=R_H^2.
\label{eq9}
\end{equation}

Therefore, these coordinates define a four-dimensional hypersurface in the
five-dimensional space. The really remarkable fact about these coordinates
is however that the de Sitter expanding metric Eq. (\ref{eq5}) on this
hypersurface is induced by the five-dimensional Minkowski metric in the
ambient~space:
\begin{equation}
ds^2=dx_0^2-dx_1^2-dx_2^2-dx_3^2-dx_4^2.
\label{eq10}
\end{equation}

This can be easily checked by explicit calculations starting from relations
Eq. (\ref{eq8}). But these coordinates are in fact based on the symmetry
considerations \cite{11}.

Thus, the de Sitter universe can be considered as a highly symmetric
four-dimensional hypersurface (a Minkowskian sphere) in circumambient
five-dimensional Minkowski space-time. Each foliation, which
endows this Minkowskian sphere with a sense of time, are on equal footing
according to the spirit of relativity. However, the cosmological circumstances
can dictate the preferred foliation. For example, the exponentially expanding
universe Eq. (\ref{eq5}) corresponds to the sense of time of comoving 
observers for which the cosmic microwave background is isotropic and the 
universe itself is spatially flat. However, other foliations exist and can 
give closed or open universes \cite{11,12}. Namely, if we define comoving FRW 
coordinates as:
\begin{eqnarray} &&
x_0=R_H\sinh{(Ht)},\;\; x_1=r\cosh{(Ht)}\sin{\theta}\cos{\phi},
\nonumber \\ &&
x_2=r\cosh{(Ht)}\sin{\theta}\sin{\phi},\;\;
x_3=r\cosh{(Ht)}\cos{\theta},\nonumber \\ &&
x_4=R_H\sqrt{1-\frac{r^2}{R_H^2}}\;\cosh{(Ht)},
\label{eq11}
\end{eqnarray}
the ambient Minkowski metric induces on the four-dimensional Minkowskian
sphere the metric:
\begin{equation}
ds^2=dt^2-\cosh^2{(Ht)}\left [\frac{dr^2}{1-\frac{r^2}{R_H^2}}
+r^2(d\theta^2+\sin^2{\theta}d\phi^2)\right ],
\label{eq12}
\end{equation}
which corresponds to the close, cosmological constant dominated, FRW universe.
While using the coordinatization
\vspace{-12pt}
\begin{eqnarray}  &&
x_0=R_H\sqrt{1+\frac{r^2}{R_H^2}}\;\sinh{(Ht)},\;\;
x_1=r\sinh{(Ht)}\sin{\theta}\cos{\phi},\nonumber \\ &&
x_2=r\sinh{(Ht)}\sin{\theta}\sin{\phi},\;\;
x_3=r\sinh{(Ht)}\cos{\theta},\nonumber \\ &&
x_4=R_H\cosh{(Ht)},
\label{eq13}
\end{eqnarray}
we get an open, cosmological constant dominated, FRW universe with the metric
\begin{equation}
ds^2=dt^2-\sinh^2{Ht}\left [\frac{dr^2}{1+\frac{r^2}{R_H^2}}
+r^2(d\theta^2+\sin^2{\theta}d\phi^2)\right ].
\label{eq14}
\end{equation}

As we see, the notion of expanding space is rather subjective. Only comoving
observers (galaxies) would find it natural, and our insistence on using this
notion reflects our belief that there is no other natural set of observers
which would define a different preferred foliation for cosmology. Nevertheless,
it seems that quasars dispute this common wisdom and do not find expanding
space natural.

Cosmological redshift is usually associated with the expanding space. However,
in contrast to the expanding space, redshift is a directly observable
objective property of light signals emitted by \mbox{distant sources.}

Suppose two light signals are emitted at cosmological times $t_e$ and
$t_e+\Delta t_e$ in a distant galaxy at the comoving coordinate $\psi$. We
at $\psi=0$ will receive signals at cosmological times $t_r$ and
$t_r+\Delta t_r$. How are $\Delta t_e$ and $\Delta t_r$ related to each other?
We can find this relation as follows \cite{6}.

Light propagates along null-geodesics or with $ds=0$ along its world-line.
Then, for radially propagating light-signals we have from Eq. (\ref{eq4})
$dt=-a(t)d\psi$ and, therefore,
\begin{equation}
\psi=\int\limits_{t_e}^{t_r}\frac{dt}{a(t)}=\int\limits_{t_e+\Delta t_e}
^{t_r+\Delta t_r}\frac{dt}{a(t)}\approx \int\limits_{t_e}^{t_r}\frac{dt}{a(t)}
+\frac{\Delta t_r}{a(t_r)}-\frac{\Delta t_e}{a(t_e)}.
\label{eq15}
\end{equation}

Hence, we obtain the cosmological time dilation relation
\begin{equation}
\frac{\Delta t_r}{\Delta t_e}=\frac{a(t_r)}{a(t_e)}.
\label{eq16}
\end{equation}

As its derivation indicates, it is an elementary and general consequence
of general relativity and space-time geometry. Usually this  time dilation
relation is applied to redshifts: if $\Delta t$ represents the period
(inverse frequency) of light-wave with the wavelength $\lambda\sim \Delta t$,
then the redshift
\begin{equation}
z=\frac{\lambda_r-\lambda_e}{\lambda_e}=\frac{a(t_r)}{a(t_e)}-1.
\label{eq17}
\end{equation}

The derivation of the cosmological time dilation relation Eq. (\ref{eq16})
is the most simple in the comoving coordinates where it can be interpreted as
a result of the space expansion. But we can still use any other
coordinates. Let us consider, for example, the de Sitter space-time in static
coordinates where nothing expands or shrinks and rederive Eq. 
(\ref{eq16}).

For radial light propagation, we have from Eq. (\ref{eq7}) that
\begin{equation}
dT=-\left (1-\frac{R^2}{R_H^2}\right )^{-1}dR.
\label{eq18}
\end{equation}
($R$ decreases when the light signal moves towards us from a distant galaxy.
Hence the minus sign). We obtain then the following relations:
\begin{equation}
T_r-T_e=-\int\limits_{R_e}^{R_r}\frac{dR}{1-\frac{R^2}{R_H^2}},\;\;
T_r+\Delta T_r-(T_e+\Delta T_e)=-\int\limits_{R_e+\Delta R_e}^{R_r+\Delta R_r}
\frac{dR}{1-\frac{R^2}{R_H^2}}.
\label{eq19}
\end{equation}

Assuming infinitesimal $\Delta T_r$ and $\Delta T_e$, we can easily get from
Eq. (\ref{eq19})
\begin{equation}
\Delta T_r-\Delta T_e=-\frac{\Delta R_r}{1-\frac{R_r^2}{R_H^2}}+
\frac{\Delta R_e}{1-\frac{R_e^2}{R_H^2}}.
\label{eq20}
\end{equation}

The coordinate $r$ of a comoving galaxy does not change with time; taking it
into account, the second relation in Eq. (\ref{eq6}) indicates that
\begin{equation}
\frac{dR}{dT}=HR\left (1-\frac{R^2}{R_H^2}\right ).
\label{eq21}
\end{equation}

Using this expression and $R_r=0$, we get from Eq. (\ref{eq20})
\begin{equation}
\frac{\Delta T_r}{\Delta T_e}=1+HR_e=1+\frac{R_e}{R_H}.
\label{eq22}
\end{equation}

But the coordinate time $\Delta T$ is not the time that clocks of the
comoving observers actually measure. The latter according to Eq. 
(\ref{eq5}) is just the cosmic time $\Delta t$, and the first equation in 
Eq. (\ref{eq6}) together with Eq. (\ref{eq21}) indicate the 
following relation between them
\begin{equation}
\frac{\Delta t}{\Delta T}=1-\frac{R/R_H}{1-\frac{R^2}{R_H^2}}\frac{dR}{dT}=
1-\frac{R^2}{R_H^2}
\label{eq23}
\end{equation}

Therefore,
\begin{equation}
\frac{\Delta t_r}{\Delta t_e}=\frac{1-R_r^2/R_H^2}
{1-R_e^2/R_H^2}\;\frac{\Delta T_r}{\Delta T_e}=\left (1-\frac{R_e}
{R_H}\right )^{-1}
\label{eq24}
\end{equation}

The fact that Eq. (\ref{eq24}) is the same relation as Eq. 
(\ref{eq16}) is not obvious, but can be shown as follows \cite{1-F}. Along 
a null radial geodesic from $(t_e,r_e)$ to  $(t_r,0)$, we have according 
to Eq. (\ref{eq5}) $dt=-e^{Ht}dr$ and respectively
\begin{equation}
r_e=-\int\limits_{r_e}^{0}dr=\int\limits_{t_e}^{t_r}e^{-Ht}dt=
\frac{1}{H}(e^{-Ht_e}-e^{-Ht_r})
\label{eq25}
\end{equation}

Then, the second equation in Eq. (\ref{eq6}) indicates that
\begin{equation}
R_e=e^{Ht_e}r_e=\frac{1}{H}(1-e^{H(t_e-t_r)})=R_H\left (1-\frac{a(t_e)}
{a(t_r)}\right )
\label{eq26}
\end{equation}
and, therefore,
\begin{equation}
\frac{a(t_e)}{a(t_r)}=\left (1-\frac{R_e}{R_H}\right )^{-1}
\label{eq27}
\end{equation}

Which proves that Eq.s (\ref{eq16}) and (\ref{eq24}) relations are 
equivalent. However, the physical interpretation of the cosmological time 
dilation in static coordinates is different. The relation following from 
Eq. (\ref{eq7}):
\begin{equation}
dt=\sqrt{\Phi(R)-\Phi^{-1}(R)\left (\frac{dR}{dT}\right )^2}\,dT
\label{eq28}
\end{equation}
between the cosmic time $t$ (the proper time of comoving observers) and the
coordinate time $T$, where
\begin{equation}
\Phi(R)=1-\frac{R^2}{R_H^2}
\label{eq29}
\end{equation}
can be interpreted \cite{1-F} as resulting from both the gravitational
redshift due to ``potential'' $\Phi(R)$, and the kinematical effect
(the Doppler shift) due to the Hubble recession velocity $\frac{dR}{dT}$
of the light-source.

As the discussion above shows, in contrast to the expanding space, which
is a subjective notion depending upon the coordinate choice, the cosmological
time dilation is an objective and very basic phenomenon with the important
observable consequence that the rate of any time variation of the
radiation emitted by a distant source with the cosmological redshift $z$
should be proportional to $(1+z)^{-1}$ (see Eq.s (\ref{eq16}) and  
(\ref{eq17})).

There is no way to escape this conclusion under the standard belief that all
distant astrophysical objects participate in the Hubble flow. Therefore, it
is not surprising that observations of distant type Ia supernova confirms
this time dilation prediction \cite{13} (it was suggested long ago to use
supernova to detect the time dilation \cite{14}).

However, a recent study of over 800 quasar light curves found no
evidence of the expected effects of time dilation \cite{15}. This is a very
surprising result with no explanation within the conventional cosmological
framework (note that some studies report the lack of time dilation signatures
also in gamma-ray burst light curves \cite{14A,14B}. See however
\cite{14C,14D}).

Some non-conventional explanations were proposed in \cite{15}, for example,
the idea that the expected effects of the quasar time dilation are exactly
compensated by an increase in the intrinsic timescale of quasar variability
due to the growth of the quasar black holes, or that the observed variations
are not intrinsic for quasars but are caused by microlensing events at much
lower redshifts. Nonetheless, these explanations do not seem plausible.

We will try to argue in this note that this surprising result, if confirmed
without any doubt, will imply a radical change of our cosmological
perspective, namely that (at least some) quasars are related to the
pre-Big-Bang objects, which we call St. Augustine's objects for reasons
explained later. These objects would define a preferred foliation different
from the preferred foliation of the Hubble flow comoving observables and this
new preferred foliation implies a pre-existing space-time predating the
Big Bang.

Perhaps some comments are appropriate here. Application of the cosmological
principle assumes that the cosmic time and the corresponding three
dimensional space, defined as a set of simultaneous events according to this
cosmic time, are well defined concepts in cosmology. It may seem that this
is obvious and should be taken for granted. However, G\"{o}del's discovery
\cite{15A} of cosmological space-time with closed timelike curves passing
through every event made it clear that in general, the existence of global
cosmic time is not at all guaranteed (the first example of space-time with
closed timelike curves was in fact given by Lanczos \cite{15AA} in 1923 and
then rediscovered by van Stockum \cite{15AAA} in 1937). The existence of
cosmic time requires a condition which is stronger than causality. Namely, a
space-time admits a cosmic time if and only if it is stably causal
\cite{15B}. Space-time is stably causal if closed timelike curves are not
only absent but they do not appear even after a deformation of the
space-time metric which slightly opens up the light cones all over the
space-time \cite{15BB,15C,15D}.

Friedmann-Robertson-Walker metric is stably causal and therefore admits
the cosmic time. However, it is the matter content of the universe (cosmic
substratum) and motion pattern of this substratum which actually endow the
cosmic time and Friedmann-Robertson-Walker metric with the real physical
meaning via celebrated Weyl's principle \cite{15E,15F}. Weyl's principle is
an assumption about the nature of the Hubble flow and it states that world
lines of the ``fundamental particles'' of the cosmic substratum (usually
assumed to be galaxies or clusters of galaxies) form, after averaging over
the peculiar motion, a space-time-filling family (congruence) of non-crossing
geodesics converging towards the common past \cite{15F}.
Friedmann-Robertson-Walker metric holds a privileged position as it is
co-moving with the fundamental observers. The above mentioned congruence of
world lines is unique in the FRW space-time and thus defines its preferred
foliation into spatial hypersurfaces that are ordered in cosmic time.

Unlike the FRW space-time, in the de Sitter space-time there is no unique
choice of congruence and that's why the de Sitter space-time can be
considered either as a static or as an expanding universe \cite{15F}.

When we speak about St. Augustine's objects it is in fact assumed that the
FRW space-time is only a part of a bigger entity (imagine, for example, an
embedding of a spatially flat FRW space-time into the five-dimensional
Minkowski space considered at the end of the paper) and that motion pattern
of the cosmic substratum in other parts of this encompassing space are quite
different from the Weyl congruence defining the Hubble flow.

\section{Other Quasar Mysteries}

Quasars (Quasi-Stellar Objects) are fascinating objects discovered half
a century ago \cite{16,17}. Shortly after their discovery, Zel'dovich
\cite{18} and Salpeter \cite{19} independently suggested accretion on a
black hole as an energy source of quasars. This hypothesis became universally
accepted after the publication of Lynden-Bell \cite{20} showing that the main
properties of quasars could indeed be understood in terms of release of
gravitational binding energy of matter accreting onto a supermassive black
hole. Nevertheless, quasars largely remain the most mysterious and enigmatic
objects in the Universe and even the most basic and major questions about
them are far from being settled \cite{21,22,23}. Let us briefly outline some
problems in quasar research with no definite answers \cite{23}.

\subsection{The Origin of Supermassive Black Holes}

If we adopt the standard view that the observed quasar redshifts are of
cosmological origin, when the unavoidable consequence is that the quasars are
immensely bright: brighter than thousands of the brightest galaxies at low
redshifts. At that strong variability of quasars in short time scales
indicates that the majority of this immense energy is emitted from a tiny
region not exceeding the Solar System in size. Supermassive black holes are
indeed the only powerhouses that can be imagined to meet such extreme energy
and size requirements.

A simple argument why the central black hole in quasars should be supermassive
goes as \mbox{follows \cite{24,25,26}.} Let us assume a steady accretion so 
that the inward gravitational force $-\rho dV \nabla \Phi$ acting on a small 
element $dV$ of accreting gas of density $\rho$ in the gravitational 
potential $\Phi$ of the black hole is balanced everywhere by the radiation 
pressure force $d\vec{F}$ on this gas element. If the gas is fully ionized 
hydrogen then radiation pressure force acts mainly on free electrons through 
Thomson scattering (for protons the scattering cross section is smaller by 
the factor $(m_e/m_p)^2\approx 2.5\cdot 10^{-8}$). However, due to Coulomb 
attraction, the electrons drag the protons with them so that the charge 
separation does not occur). The Thomson cross section $\sigma_T$ gives the 
effective area for an electron to absorb radiation when it is illuminated. 
Therefore, if the radiation energy flux is $d\vec{\Sigma}$, each electron 
absorbs momentum at a rate $(d\vec{\Sigma}/c)\sigma_T$. The radiation pressure 
force $d\vec{F}$ equals the rate at which the volume element $dV$ absorbs 
momentum. In this volume element we have $N=\rho dV/m_p$ protons and hence, 
the same amount of electrons, as the gas is assumed to be fully ionized. 
Therefore, $$d\vec{F}=N\frac{d\vec{\Sigma}}{c}\sigma_T=\frac{\rho dV}{m_p}\,
\frac{d\vec{\Sigma}}{c}\sigma_T$$

Then the condition of hydrostatic equilibrium, $-\rho dV \nabla \Phi+d\vec{F}=
0$, gives the equation $$d\vec{\Sigma}=\frac{m_pc}{\sigma_T}\; \nabla \Phi$$

Using this relation, the Gauss's theorem and the Poisson equation
$\nabla^2 \Phi=4\pi G\rho$, $G$ being the Newton constant, we can calculate
the total luminosity as a surface integral over some closed surface $S$ around
the quasar:
$$L=\int_S d\vec{\Sigma}\cdot d\vec{S}=\frac{m_pc}{\sigma_T}\int_S \nabla \Phi
\cdot d\vec{S}=\frac{m_pc}{\sigma_T}\int_V \nabla^2 \Phi
dV=4\pi G\,\frac{m_pc}{\sigma_T}\int_V\rho dV$$

Therefore, finally,
\begin{equation}
L_E=\frac{4\pi Gm_pMc}{\sigma_T}
\label{eq30}
\end{equation}
where $M$ is the quasar mass. This so called Eddington luminosity is an
estimate of the maximal luminosity a quasar of mass $M$ can have if it is
powered by a steady accretion. Substituting \mbox{$\sigma_T\approx 6.7\cdot 
10^{-25}~\mathrm{cm}^2$} and other physical constants, Eq. (\ref{eq30}) 
can be rewritten as follows \cite{24}
\begin{equation}
L_E\approx 1.3\cdot 10^{38}\left(\frac{M}{M_\odot}\right)~\mathrm{erg\;s}^{-1}
\label{eq31}
\end{equation}
where $M_\odot$ is the solar mass. Typical quasar luminosity is
$L_Q\approx 10^{46}~\mathrm{erg\; s}^{-1}$ \cite{26}. Then Eq. 
(\ref{eq31})indicates that a central mass of the order of $10^8M_\odot$ is 
required.

Various methods were developed in past decades to estimate
quasar masses from observations \cite{27} and the results confirm the above
conclusion that quasars are supermassive. Moreover, the results indicate
\cite{27} that quasars with masses $10^9M_\odot$ were probably already present
at $z\sim 7$ when the universe was less than 1 Gyr old. This raises
a potential problem: how these supermassive black holes have had grown up
given the limited time they had?

If the accretion rate is $\mu$, then the associated luminosity can be written
as $L=\epsilon\mu c^2$, with $\epsilon$ representing the radiative efficiency
that equals the fraction of the accreted rest energy converted into radiation.
The growth rate of the black hole under such accretion is
$\dot{M}=(1-\epsilon)\mu$. Therefore, from these two relations, using the
Eddington luminosity Eq. (\ref{eq30}), we get
\begin{equation}
\dot{M}=\frac{1-\epsilon}{\epsilon}\,\frac{L_E}{c^2}=
\frac{1-\epsilon}{\epsilon}\,\frac{M}{t_S}
\label{eq32}
\end{equation}
where
\begin{equation}
t_S=\frac{\sigma_T\, c}{4\pi G m_p}\approx 4.5\cdot 10^8~\mathrm{yr}
\label{eq33}
\end{equation}
is the so called Salpeter time. Equation (\ref{eq32}) indicates that under
the steady Eddington accretion the black hole mass grows exponentially with
time:
\begin{equation}
M=M_0\exp{\left(\frac{1-\epsilon}{\epsilon}\,\frac{(t-t_0)}{t_S}\right)}
\label{eq34}
\end{equation}
where $M_0$ is the mass of the black hole seed. What can we say about the
radiative efficiency $\epsilon$? To estimate $\epsilon$, we can use a
Newtonian argument. In fact not quite Newtonian, because the pure Newtonian
estimate, assuming that the accretion disk extends up to the Schwarzschild
radius, is a factor of two higher than the more appropriate general
relativistic result. However, it was shown be Paczy\'{n}sky and Wiita that
the following simple pseudo-Newtonian potential
\begin{equation}
\Phi=-\frac{GM}{r-r_S},\;\;\;r_S=\frac{2GM}{c^2}
\label{eq35}
\end{equation}
reproduces the correct general relativistic results surprisingly accurately
as far as the black hole accretion flaw is concerned \cite{28,29}. Here,
$r_S$ is the Schwarzschild radius of the black hole.

For a test body of mass $m$ moving on a circular orbit of radius $r$ in the
potential Eq. (\ref{eq35}), we find for its velocity
$$V=\frac{\sqrt{GMr}}{r-r_S}$$
and for its total energy
\begin{equation}
E=\frac{mV^2}{2}+m\Phi=-GMm\frac{r-2r_S}{2(r-r_S)^2}
\label{eq36}
\end{equation}

As we see, only for $r>2r_S$ the bound circular orbits are possible. But not
all of them are stable. The stability condition is the following:

$$\frac{d^2U}{dr^2}=\frac{d^2}{dr^2}\left (m\Phi+\frac{l^2}{2mr^2}\right )=
m\frac{d^2\Phi}{dr^2}+3\frac{l^2}{mr^4}>0$$
where $U$ is the effective potential corresponding to the orbital momentum
$l$. The orbital radius $r$ is determined from the condition
$$\frac{dU}{dr}=m\frac{d\Phi}{dr}-\frac{l^2}{mr^3}=0$$

This enables us to exclude $l$ and rewrite the stability condition in the form
$$m\left(\frac{d^2\Phi}{dr^2}+\frac{3}{r}\frac{d\Phi}{dr}\right )=
\frac{GMm}{r(r-r_S)^3}\,(r-3r_S)>0$$

Therefore, only circular orbits with $r>3r_S$ are stable and hence, the steady
accretion disk ends at $r=3r_S$. The binding energy at this innermost stable
orbit is, according to Eq. (\ref{eq36}), $$-E=\frac{GMm}{8r_S}=
\frac{1}{16}mc^2$$

When the particle of mass $m$ migrates from the outskirts of the accretion
disk at $r\gg r_S$, where its binding  energy $-E\approx 0$, to the inner
edge of the accretion disk, this $mc^2/16$ energy is released as heat and
eventually radiation. Therefore, the radiative efficiency of such a
Schwarzschild black hole engine is $\epsilon=1/16\approx 6.2\%$. More correct
treatment of general relativistic effects gives $\epsilon\approx 5.7\%$
\cite{30}. Rotating black holes can lead to even higher
radiative efficiencies: up to about $30\%$ in realistic case \cite{30}.
For comparison, nuclear burning of hydrogen to helium in stars has the
efficiency of only $0.7\%$.

Usually, in quasar research one assumes a canonical radiative efficiency
of $\epsilon=0.1$ \cite{27}. For \mbox{$M=10^9M_\odot$} at $z\sim 7$ and
$t-t_0\sim 0.7$~Gyr that corresponds to $z_0\sim 30$ when the first stars were
formed, we get from Eq. (\ref{eq34}) that the black hole seeds of mass
$M_0\sim 10^3M_\odot$ is needed. This seems somewhat high as it is
expected that the black hole remnants from the first generation stars had
typical values less than tens of solar masses \cite{27}.

Although this is not considered as an immediate problem, because we could
imagine ways how to produce such massive black hole seeds (for example,
direct collapse of primordial gas clouds \cite{27}), but it will be fair to
say that we do not completely understand the origin of supermassive black
holes at present. Even if we would have a sufficiently massive black hole
seed, it is not clear how a steady accretion at Eddington rates can be
maintained for such a long period of time. The accreting gas has to lose
about $99.9\%$ of its angular momentum to migrate from the host galaxy to
the vicinity of the black hole \cite{31}. During this time, various processes
can intervene to make the accretion not continuous. For example, star
formation which inevitably occur in the accretion disks on large scales, can
reduce the amount of gas which can be directly accreted onto the black hole
\cite{24,31}. In fact, our present day arguments about how quasars ignite
and produce their stupendous energies are of the kind of hand-waving lacking
necessary details \cite{32} and so we still do not have good predictive
physical models \cite{22}.

\subsection{Evolution of Quasars and Age Problems}

Another mysterious thing about quasars which is poorly understood is the
evolution of quasar luminosity over redshifts: quasars at high redshifts
are far more luminous than quasars at low \mbox{redshifts \cite{23}.} But, 
strangely, quasar spectra show little evidence, if any, of this evolution:  
quasar spectral features at low and high redshifts are very similar \cite{23}.
It seems that quasars were already mature at $z\sim 6-7$
and the physical properties of their accretion disks have not changed since
then despite strong evolution of their luminosities.

Not only quasars at high redshifts seem to be mature but their surroundings
too. It is assumed that it requires many generations of supernovae to
produce sizable amount of metals. So, the natural expectation is that the
accreting material for high redshift quasars should be metal-free.
Observations, on the contrary, indicate that metallicities are even higher
than the solar metallicity \cite{23}. Recent discovery of $z=7.51$ galaxy
with significant metallicity and very high star formation rate \cite{33}
also indicates more rapid chemical evolution in the early universe than it
is usually assumed.

Some quasars were found \cite{34} at $z\sim 6$ without any detectable infrared
emission from the hot dust which is expected in the standard accretion disk
model and is indeed universally present at low redshifts. This may be a sign
of the long evolution of quasars absent in the ultraviolet, optical and
X-ray bands of their spectra. However, the rarity of such dust-free quasars
indicate a very rapid mass accretion and dust formation at their alleged early
evolutionary stages \cite{34}.

Another question is what ignites a quasar. It is usually assumed that
the major merger events of gas-rich galaxies trigger the accretion on the
central black hole and ignite a quasar. However about half of host galaxies
have undistorted disks and show no signs of distortions induced by merger
\cite{23}.

Besides the problem of explaining the appearance at a surprisingly early
times after the Big Bang of chemically mature galaxies hosting a supermassive
black holes, there is still another age problem related to quasars. Namely,
a quasar was discovered at $z=3.91$ whose age is 2--3~Gyr, while according to
the standard cosmological model the Universe was only 1.63~Gyr old at this
redshift \cite{35}. Therefore, this quasar appears to be older than the
Universe---an obvious paradox requiring explanation.

While the so called $\alpha$-process elements ($O$, $Ne$, $Mg$, $Si$, $S$) are
mainly produced in type II supernovae, the so called iron-peak elements ($V$,
$Cr$, $Mn$, $Fe$, $Co$, $Ni$) are produced in type Ia supernovae \cite{36,37}.
The corresponding characteristic time scales are vastly different which
enables to estimate the age of astrophysical objects by measuring their
$Fe/\alpha$ abundance ratios.

Type II supernovae are explosions of massive stars induced by the collapse of
their inert iron-nickel cores when the core exceeds the Chandrasek\-har limit
of about 1.4 solar masses and the electron degeneracy pressure in the core is
no longer sufficient to maintain the equilibrium against gravity. The lifetime
of such massive stars ($M>8M_\odot$) is quite small, about $10^6-10^7$ years.

Type Ia supernovae are thermonuclear explosions of accreting carbon-oxygen
white dwarfs in close binaries. The lifetime of such binaries, estimated from
the chemical evolution of our galaxy, is about 1 Gyr \cite{37}, although
a lifetime distribution can span within a wide range of 0.1--20 Gyr \cite{38}.

As it was already mentioned, this difference in timescales gives us a specific
``clock'' to measure the age of astrophysical objects. In the case of the
quasar APM 08279+5255 at $z=3.91$, using the measured $Fe/O$ abundance ratio
and a detailed chemodynamical model, it was estimated \cite{39} that the age
of this quasar is $t_Q=2.1\pm 0.3$~Gyr.

The age of the Universe at this redshift can be obtained in the following way.
From Eq. (\ref{eq17}), with the identifications $t_e=t$, $t_r=t_0$, 
where $t_0$ is the present age, we get
\begin{equation}
\frac{dz}{dt}=-\frac{a(t_0)}{a^2(t)}\,\dot{a}(t)=-(1+z)H(t)
\label{eq37}
\end{equation}

On the other hand, the Hubble parameter $H=\dot{a}/a$ is given by the Friedmann
equation which for the flat Universe ($k=0$) takes the form (see, for example,
\cite{6}):
\begin{equation}
H^2=\frac{8\pi G}{3}(\rho_m+\rho_r+\rho_v)
\label{eq38}
\end{equation}
$\rho_m$, $\rho_r$ and $\rho_v$ being the energy densities of matter,
radiation and vacuum, respectively. The vacuum energy density $\rho_v$ (the so
called dark energy) does not depend on the scale factor. The matter energy
density $\rho_m$ scales as $a^{-3}$ and the radiation energy density $\rho_r$
as $a^{-4}$ (because in this case not only the photon number density
decreases as $a^{-3}$ upon space expansion, but also each photon experiences
a redshift and its energy decreases as $a^{-1}$). Therefore, we get from
Eq. (\ref{eq38}),
\begin{equation}
\frac{H^2}{H_0^2}=\Omega_{m0}(1+z)^3+\Omega_{r0}(1+z)^4+1-\Omega_{m0}-
\Omega_{r0}
\label{eq39}
\end{equation}
where
$$\Omega_{m0}=\frac{\rho_{m0}}{\rho_{c0}},\;\;\;
\Omega_{r0}=\frac{\rho_{r0}}{\rho_{c0}}$$
are the present fractional energy densities of matter and radiation given
in terms of the critical density
$$\rho_{c0}=\frac{3H_0^2}{8\pi G}$$
and we have taken into account that for the flat Universe $\rho_{m0}+\rho_{r0}
+\rho_{v0}=\rho_{c0}$.

Equation (\ref{eq39}) gives the Hubble parameter as a function of redshift.
Therefore, we can separate variables in Eq. (\ref{eq37}) and get the 
following integral (note that at the Big Bang, \textit{t} = 0, the scale 
factor is zero and hence the redshift is infinity) \cite{39}:
\begin{eqnarray} &&
t=-\int\limits_\infty^z\frac{dz}{(1+z)H(z)}=
\frac{1}{H_0}
\int\limits_z^\infty\frac{dz}{(1+z)\sqrt{\Omega_{m0}(1+z)^3+
\Omega_{r0}(1+z)^4+1-\Omega_{m0}-\Omega_{r0}}}
\label{eq40}
\end{eqnarray}

This is an elliptic integral. However, if we ignore the radiation energy
density, which is very small at present epoch ($\Omega_{r0}<10^{-4}$) and
becomes important only at very high redshifts hence contributing little in the
integral Eq. (\ref{eq40}), then the integral can be calculated in terms of
elementary functions by a~sequence of substitutions \cite{40}
$$x=\frac{1}{1+z},\;\;\;\xi=x^3,\;\;\;\eta=\sqrt{(\Omega_{m0}^{-1}-1)\xi}$$
and we finally get \cite{40,41}
\begin{equation}
t=\frac{2}{3H_0\sqrt{1-\Omega_{m0}}}
\operatorname{arsinh}{\sqrt{\frac{\Omega_{m0}^{-1}-1}{(1+z)^3}}}
\label{eq41}
\end{equation}

For $z=3.91$, $H_0= 70.4~\mathrm{km}\mathrm{s}^{-1}\mathrm{Mpc}^{-1}$ and
$\Omega_{m0}= 0.27$,  Eq. (\ref{eq41}) gives $t\approx 1.63$~Gyr and we
have the above mentioned age problem $t<t_Q$.

\subsection{Correlation of Quasars with Galaxies of Lower Redshifts}

A long standing controversy is an alleged association of high-redshift quasars
with low-redshift galaxies \cite{23,42,43,44}. Some quasars are surprisingly
close, only a few arcseconds away, to the galaxies with widely different
redshift. In some amazing cases filaments apparently connecting the
pairs of objects with different redshifts have been observed. Maybe one of the
most impressive cases is the Seyfert galaxy NGC 7603 \cite{45,46}. Seyfert
galaxies, less energetic cousins of quasars, are spiral or irregular galaxies
containing an extremely bright quasar-like nucleus. The redshift of NGC 7603
is $z=0.029$. There is another smaller galaxy, NGC 7603B, with $z=0.057$
nearby and the two are apparently connected by a narrow luminous filament,
NGC 7603B lying at the end of the filament. The redshift of the filament
itself is $z=0.030$ (the same as for NGC 7603 within the measurement errors
which are of the order of $0.001$ \cite{45}). There are two more galaxies
lying exactly on the filament: one with the redshift $z=0.243$ positioned
near NGC 7603B, and the another one with $z=0.391$ positioned near NGC 7603
\cite{45,46} (see Figure 1a in \cite{45}).

Mainstream cosmologists simply ignore this kind of evidence supporting
association of objects with different redshifts considering all these cases
as a just random projections of background objects like constellations. They
might be right in some cases but the statistics of anomalies seems to be too
high and indicates that we do not completely understand these phenomena
\cite{23}.

\subsection{Apparent Superluminal Motion}

The last conundrum related to quasars which we will discuss in this
section is the problem of apparent superluminal motion \cite{23,43,48A,48}.
In fact this phenomenon was predicted \cite{47} before it was discovered
following the development of Very Large Base Interferometry (VLBI) that
allowed a very accurate mapping of the radio sources morphologies. It was
found that in many radio galaxies and quasars there are several compact
sources of radio emission that appear to move apart superluminaly in
successive VLBI images \cite{49}.

Canonical explanation of this apparent superluminal motion (see, for example,
\cite{26}) is ingenious but simple enough to be used as an undergraduate
problem for the first year physics students \cite{50}. The standard
derivation, however, is valid only for nearby sources as it ignores
cosmological space expansion, and this fact is not sufficiently stressed in
existing literature \cite{51}. Therefore, to avoid a confusion, we provide,
following \cite{51}, a derivation valid for all redshifts.

\begin{figure}[ht]
 \begin{center}
    \includegraphics{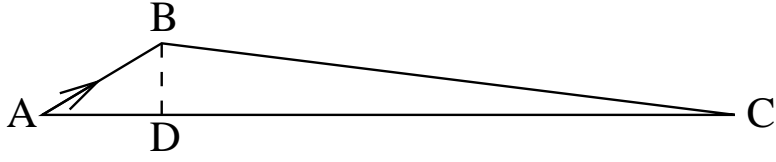}
  \end{center}
\caption{Relativistic beaming model of apparent superluminal motion}
\label{fig2}
\end{figure}
Suppose two radio emitting blobs coincide at a point $A$ (see Figure 
\ref{fig2}) at cosmic time $t_e$. The observer at a point $C$ notices this 
coincidence at a later time $t_r$ because of the finite speed of light. One 
blob remains stationary while the other moves at a relativistic speed $V$ 
towards the observer under a small angle $\theta=\angle BAC$ to the 
line-of-sight of the
observer. After infinitesimal time $\Delta t_e$, the moving blob will reach
the point B. The observer notices this event at time $t_r+\Delta t_r$
with some $\Delta t_r$. As was already explained, in FRW universe the
emission and reception times along radial null-geodesics are related as follows
(in units where $c=1$, so below we will use $V$ and $\beta=V/c$
interchangeably)
\begin{equation}
\psi=\int\limits_{t_e}^{t_r}\frac{dt}{a(t)},\;\;\;
\psi+\Delta\psi=\int\limits_{t_e+\Delta t_e}^{t_r+\Delta t_r}
\frac{dt}{a(t)}\approx \int\limits_{t_e}^{t_r}\frac{dt}{a(t)}-\frac{\Delta
t_e}{a(t_e)}+\frac{\Delta t_r}{a(t_r)}
\label{eq42}
\end{equation}
where $\psi$ is the co-moving radial coordinate of the point $A$, $\psi+
\Delta\psi$---of the point $B$ and it is assumed that the observer is
at the origin $\psi=0$. From Eq. (\ref{eq42}) we have
\begin{equation}
\Delta \psi=\frac{\Delta t_r}{a(t_r)}-\frac{\Delta t_e}{a(t_e)}
\label{eq43}
\end{equation}

On the other hand, the point $B$ has, at the first order in $\Delta t_e$,
the same radial coordinate as the point $D$. Therefore,
\begin{equation}
a(t_e)\Delta\psi\approx -AD=-V\cos{\theta}\,\Delta t_e
\label{eq44}
\end{equation}

Combining Eq.s (\ref{eq43}) and (\ref{eq44}) we get, after using 
Eq. (\ref{eq17}), $$\Delta t_r=(1+z)(1-\beta\cos{\theta})\Delta t_e$$

Therefore, the observed apparent transverse velocity of the moving blob with
respect to the stationary one is \cite{51}
\begin{equation}
V_a=\frac{BD}{\Delta t_r}=\frac{V\sin{\theta}\,\Delta t_e}{\Delta t_r}=
\frac{V\sin{\theta}}{(1+z)(1-\beta\cos{\theta})}
\label{eq45}
\end{equation}

The quantity which is actually measured on Earth is the observed angular
proper motion of the moving blob on the sky per time unit $\mu=\Delta\alpha/
\Delta t_r$ and to transform it into the apparent transverse velocity
$V_a=\mu D_A$ we need to know the so called angular diameter distance $D_A$.
It is defined in such a way \cite{6} that for a small solid angle $\Delta
\Omega=2\pi(1-\cos{(\Delta\alpha)})\approx \pi (\Delta\alpha)^2$ the
corresponding cross-sectional area $\Delta S$ at a distance $D_A$ is
\begin{equation}
\Delta S=D_A^2\Delta\Omega\approx\pi D_A^2(\Delta \alpha)^2
\label{eq46}
\end{equation}

On the other hand, for the FRW metric Eq. (\ref{eq4}), the total area 
of a sphere
$\psi=\mathrm{const}$ at cosmic time $t_e$ is $4\pi a^2(t_e)S^2(\psi;k)$ and
this area corresponds to the solid angle $4\pi$. Therefore,
\begin{equation}
\Delta S=\frac{\Delta\Omega}{4\pi}\,4\pi a^2(t_e)S^2(\psi;k)\approx
\pi (\Delta\alpha)^2a^2(t_e)S^2(\psi;k)
\label{eq47}
\end{equation}
From Eq.s (\ref{eq46}) and  (\ref{eq47}), we get $D_A=a(t_e)S(\psi;k)$. In
particular, for the spatially flat universe,
\begin{equation}
D_A=a(t_e)\psi
\label{eq48}
\end{equation}

To express Eq. (\ref{eq48}) in terms of observed quantities, we note 
again that for radial null-geodesics $d\psi=-dt/a(t)$ and therefore, by using
Eq.s (\ref{eq17}) and (\ref{eq39}), we get \cite{51}
\begin{eqnarray} &&
D_A=\frac{a(t_e)}{a_0}\int\limits_{t_e}^{t_0}\frac{dt}{a(t)/a_0}=
\frac{1}{1+z}\int\limits_0^z\frac{dz^\prime}{H(z^\prime)}= \nonumber \\ &&
\frac{1}{(1+z)H_0}\int\limits_0^z\frac{dz^\prime}{\sqrt{\Omega_{m0}
(1+z^\prime)^3+1-\Omega_{m0}}}\nonumber
\end{eqnarray}

As we see, the apparent transverse velocity can be calculated from the
observed quantities as follows
\begin{equation}
\beta_a=\frac{\mu}{(1+z)H_0}\int\limits_0^z\frac{dz^\prime}{\sqrt{\Omega_{m0}
(1+z^\prime)^3+1-\Omega_{m0}}}
\label{eq49}
\end{equation}

To concentrate on the intrinsic parameters of the source, it is a common
practice to eliminate the redshift dependence of the apparent transverse
velocity by considering (we have recovered the light velocity $c$ for
a moment)

\begin{equation}
{\bar{V}}_a=\frac{c\mu}{H_0}\int\limits_0^z\frac{dz^\prime}{\sqrt{\Omega_{m0}
(1+z^\prime)^3+1-\Omega_{m0}}}=\frac{V\sin{\theta}}{1-\beta\cos{\theta}}
\label{eq50}
\end{equation}

Rather misleadingly, ${\bar{V}}_a$ is widely called ``apparent velocity''
despite the fact that it is the apparent velocity only for a co-moving
observer near the source \cite{51}. The true apparent velocity for
a terrestrial observer at a cosmological distance from the source is $V_a$,
as our derivation above indicates, and it is suppressed by a factor $1+z$
\cite{51}.

As a function of the angle $\theta$, ${\bar{V}}_a$ has a maximum
${\bar{V}}_{am}=\gamma V$ when $\theta=\theta_m$, where $\cos{\theta_m}=
\beta$. As we see, ${\bar{V}}_{am}$ can be arbitrarily large and at first
sight relativistic beaming model gives a natural solution of the apparent
superluminal motion problem. However, matters are not as simple as they seem.

We need $\gamma\gg 1$, if we want ${\bar{V}}_{am}\gg c$. But then $\beta
\sim 1$ and $\theta_m$ is very small. If quasar jets are isotropically
distributed in space, then the probability of observing an approaching jet
with $\theta<10^\circ$ is (assuming that the quasar jets are double-sided and
symmetrical) $p=2\pi(1-\cos{\theta})/(2\pi)\approx \theta^2/2\approx 0.015$.
This quite small probability indicates that the observation by chance of an
apparent superluminal motion in quasars should be relatively rare phenomenon.
It was clear from the beginning that real observations do not confirm this
conclusion \cite{52}. In fact apparent superluminal motion is common and
ubiquitous in quasars \cite{43,53}. For example, the MOJAVE (MOnitoring of
Jets in AGN with VLBA Experiments) survey contains 101 quasars with 354
observed radio components, 95\% of which move with apparent superluminal
velocities with respect to the core. The maximum observed velocity equals
to $60c$ and about 45\% of radio emitting blobs have velocities larger than
$10c$ \cite{53}.

A close inspection of the formula (\ref{eq50}) indicates that it requires
roughly $\theta<10^\circ$ to obtain ${\bar{V}}_a>10c$. Then randomly, using
the above mentioned probability $p\approx 0.015$, one would expect only
$354*p\approx 5$ blobs, out of 354, to have ${\bar{V}}_a>10c$. In reality
the observed number is 158---a totally improbable event by chance \cite{53}.

At this point, the relativistic beaming model is clearly in trouble.
Fortunately for the standard paradigm, the Doppler boosting which is
intrinsically related to the relativistic beaming comes at rescue. In
particular, the observed flux density, $F_\nu$, from an emitting blob which
moves with a relativistic velocity towards observer making an angle $\theta$
with the line of sight is enhanced compared to the intrinsic flux from the
blob at rest, $F_{0,\nu}$, according to the formula \cite{52,54}
\begin{equation}
F_\nu=\frac{F_{0,\nu}}{[\gamma(1-\beta\cos{\theta})]^{3+\alpha}}\;
\label{eq51}
\end{equation}
where $\alpha$ is the spectral index of the radiating source (in other words,
it is assumed that $F_\nu\sim \nu^{-\alpha}$). This relation follows from the
relativistic invariance of $I_\nu/\nu^3$ and the Doppler blue-shift
formula for the frequency $\nu_0=\gamma(1-\beta\cos{\theta})\nu$ \cite{55}.
Here $I_\nu$ is the specific intensity related to the flux density as follows
$F_\nu=\int I_\nu\cos{\theta^\prime}\,d\Omega^\prime$ \cite{56}. Indeed,
since $I_\nu/\nu^3=I_{0,\nu_0}/\nu_0^3$, we have
$$F_\nu=\int\left(\frac{\nu}{\nu_0}\right )^3I_{0,\nu_0}\cos{\theta^\prime}\,
d\Omega^\prime=\left(\frac{\nu}{\nu_0}\right )^3F_{0,\nu_0}$$

But ($k_0$ is some constant)
$$F_{0,\nu_0}=k_0\nu_0^{-\alpha}=\left(\frac{\nu}{\nu_0}\right )^\alpha
k_0\nu^{-\alpha}=\left(\frac{\nu}{\nu_0}\right )^\alpha F_{0,\nu}\;$$
and, therefore,
$$F_\nu=\left(\frac{\nu}{\nu_0}\right )^{3+\alpha}F_{0,\nu}$$
which is equivalent to Eq. (\ref{eq51}).

To prove that $I_\nu/\nu^3$ is indeed relativistic invariant, we first show,
following \cite{56} (see also \cite{56L}), that a phase space volume element
$d^3r\,d^3p$ is Lorentz invariant. Let a group of particles occupy an
infinitesimal volume $d^3r$ around some radius-vector and have a slight
spread $d\vec{p}$ of momenta around a mean momentum in some inertial
reference frame $S$. In the frame $S^\prime$, comoving with the particles,
the mean momentum equals to zero and all particles have the same energy
$mc^2$ to the first order because the infinitesimally small momenta of
individual particles contribute quadratically to their energy and therefore
this contribution vanishes to the first order. Let us imagine some particle
from the group with a momentum $\vec{p}^\prime$ in the frame $S^\prime$.
Then in the frame $S$ we have
$$p_x=\gamma(p_x^\prime+\beta mc),\;\;\; p_y=p_y^\prime,\;\;\;
p_z=p_z^\prime$$
and therefore the momentum spreads in the frames $S$ and $S^\prime$ are
related as follows
\begin{equation}
dp_x=\gamma\,dp_x^\prime,\;\;\; dp_y=dp_y^\prime,\;\;\; dp_z=dp_z^\prime
\label{eq52}
\end{equation}

On the other hand, due to Lorentz contraction,
\begin{equation}
dx=\frac{dx^\prime}{\gamma},\;\;\; dy=dy^\prime,\;\;\; dz=dz^\prime
\label{eq53}
\end{equation}

There is one subtlety here which sometimes can lead to confusion (see, for
example, \cite{56A,56B}). We need particles to occupy a given volume $d^3r$
element simultaneously. But simultaneity is relative: if a group of
particles are all inside a volume element $d^3r$ at time $t$ in reference
frame $S$, in the comoving frame $S^\prime$ they are inside the corresponding
volume element $d^3r^\prime=d^3r/\gamma$ at different times and according to
the Lorentz transformation the spread of these times around the mean time
$t^\prime$ is $-\gamma\beta dx/c$. In the frame $S^\prime$ the particles are
not in general at rest but move (in the $x$-direction) with infinitesimal
velocities of the order of $dp_x^\prime/m=dp_x/(m\gamma)$. Therefore, to
bring all particles at the time instant $t^\prime$ will require a shift
of their $x$-coordinates of the order of
\begin{equation}
\Delta x=\gamma\beta \frac{dx}{c}\,
\frac{dp_x}{m\gamma}=\frac{\beta dx\,dp_x}{mc}
\label{eq54}
\end{equation}

Analogously, the required shifts in the $y$ and $z$ directions are
\begin{equation}
\Delta y=\frac{\beta \gamma dx\,dp_y}{mc},\;\;\;\Delta z=\frac{\beta \gamma
dx\,dp_z}{mc}
\label{eq55}
\end{equation}

Fortunately, for finite $m$ and $\gamma$, these quantities are of the second
order and hence, can safely be ignored. Let us stress, however, that for this
argument to be valid it is essential for $S^\prime$ to be the  instantaneous
comoving frame because only in this frame particle velocities are
infinitesimal. If $S^\prime$ is not the comoving frame, we can no longer in
general ignore the effects of desynchronization, and if we do, we get an
erroneous conclusion, as in \cite{56A}, that the phase space volume element
is not Lorentz~invariant.

Relations Eq.s (\ref{eq52}) and (\ref{eq53}) do indeed show that
\begin{equation}
dx\,dy\,dz\,dp_x\,dp_y\,dp_z=dx^\prime\,dy^\prime\,dz^\prime\,dp_x^\prime\,
dp_y^\prime\,dp_z^\prime
\label{eq56}
\end{equation}

Of course this reasoning is strictly valid only for particles of finite mass.
However, the final result Eq. (\ref{eq56}) has no reference altogether 
to the particle mass and it is reasonable to propose that it should be true 
also in the $m\to 0$ limiting case of photons. Note, however, that Eq.s 
(\ref{eq54}) and (\ref{eq55}) indicate that the $m\to 0$ limit is rather 
subtle. More formal and rigorous derivations of Eq. (\ref{eq56}) for 
photons can be found in \cite{57,58}.

Now let us consider an energy content $dE$ of a phase space volume element
$d^3r\,d^3p$ occupied by $dN$ monochromatic photons of frequency $\nu$. It can
be calculated in two different ways. First, from the definition of specific
intensity, we have $dE=I_\nu d\nu\,d\Omega\,dt\,dA$, where $dA$ is the
cross-sectional area of $d^3r$ in the direction perpendicular to the photons
mean momentum. {On the other hand,} $$dE=h\nu\,dN=h\nu f d^3r\,d^3p=h\nu f\,
cdt\,dA\,p^2dp\,d\Omega=\frac{h^4\nu^3}{c^2}f\,d\nu\,d\Omega\,dt\,dA$$
where we have used the relation $d^3r=cdt\,dA$ and $f=\frac{dN}{d^3r\,d^3p}$
is the photon distribution function which is Lorentz-invariant because both
$dN$ and $d^3r\,d^3p$ are Lorentz-invariant. Comparing these two expressions
for $dE$,{ we see that} \cite{58} $$I_\nu=\frac{h^4\nu^3}{c^2}f$$
This completes the proof that $I_\nu/\nu^3$ is  Lorentz-invariant.

According to Eq. (\ref{eq51}), relativistic beaming has a very strong 
effect on the observed luminosity. For $\gamma\sim 10$, $\alpha\sim 0$ and 
$\theta\sim 10^\circ$, the enhancement is about two orders of magnitude 
(this is true
for a simple ballistic model of one radiating approaching blob for which
Eq. (\ref{eq51}) applies. In the case of more realistic jet geometries the
enhancement factor may be smaller \cite{59}). This Doppler boosting effect
allows to explain the above mentioned huge discrepancy in the statistical
properties of observed superluminal sources as a kind of the Malmquist bias
\cite{53}. Simply, due to the Doppler boosting, it is natural to expect that
flux-limited surveys of radio emitting sources will indeed preferentially
pick up brighter sources that are ejecting material with small $\theta$.
However, this argument assumes that the main part of the observed flux, even
the one that comes from the core, originates from relativistic jets, and not
everybody is ready to except this assumption \cite{60}.

To conclude this section, at present we have no compelling reason to reject
the standard cosmological concepts about quasars. However, it seems that there
is a growing number of alleged observational facts which are difficult
to reconcile with the standard paradigm. It is not a good strategy to simply
ignore these facts, in our opinion. It would be much better to remember
the nineteenth century humorist  Josh Billings' aphorism  ``It ain't ignorance
causes so much trouble; it's folks knowing so much that \mbox{ain't so''
\cite{JB}} and keep eyes wide open to new insights.

\section{The Milne Model}

Let us consider a FRW universe with vanishing total energy density $\rho$ and
vanishing cosmological constant $\Lambda$. The Friedmann equation \cite{6}
$$H^2\equiv \left (\frac{\dot{a}}{a}\right )^2=\frac{8\pi G}{3}\;\rho+
\frac{\Lambda a^2}{3}-\frac{k}{a^2}$$
then gives (for not static universe) $k<0$ and $\dot{a}=\sqrt{-k}$.
By rescaling the radial coordinate and time, we may assume that the units are
such that the curvature constant is $k=-1$ together with $c=1$. Therefore, in
this case the scale factor $a(t)=t$ linearly grows with time. The possible
integration constant can always be eliminated by a suitable change of the time
origin. The metric Eq. (\ref{eq4}) takes the form
\begin{equation}
ds^2=dt^2-t^2\left [d\psi^2+\sinh^2{\psi}\;(d\theta^2+\sin^2{\theta}\;
d\phi^2)\right ]
\label{eq57}
\end{equation}

Such a space-time is called the Milne universe. It was Milne who realized
that such a model must have an alternative, special
relativistic description because it has the gravity ``turned off''
(vanishing energy density) and no cosmological constant \cite{61,62,63,64}.
Indeed, if we introduce new coordinates
\begin{equation}
T=t\,\cosh{\psi},\;\;\; R=t\,\sinh{\psi}
\label{eq58}
\end{equation}
the metric Eq. (\ref{eq57}) takes the form
$$ds^2=dT^2-dR^2-R^2(d\theta^2+\sin^2{\theta}\;d\phi^2)$$
which is just the Minkowski line element in spherical coordinates. Let us
note that the Milne coordinates $t$ and $\psi$ cover only a part of the
Minkowski space-time. Indeed, equations Eq. (\ref{eq58}) indicate that 
$T>0$ (because the cosmic time $t>0$, $t=0$ corresponding to the Big Bang)
and $R<T$. Therefore, the Milne universe corresponds to the future light cone
of the Big Bang event. To cover the past light cone, we can redefine the Milne
coordinates $\psi\ge 0$, $t\ge 0$ in such a way that
\begin{equation}
T=-t\,\cosh{\psi},\;\;\; R=t\,\sinh{\psi}
\label{eq59}
\end{equation}

This leads to the same metric Eq. (\ref{eq57}) with a singularity at 
$t=0$. For the expanding Milne universe Eq. (\ref{eq58}), the singularity
corresponds to the Big Bang, and for the contracting Milne universe (with
$-t\le 0$ as a cosmic time), the singularity corresponds to the Big Crunch.

Equation (\ref{eq58}) give $R/T=\tanh{\psi}$, which for constant $\psi$
represents a world-line of an observer uniformly moving radially with rapidity
$\psi$. This clarifies the physical meaning of the Rindler coordinate $\psi$
and suggests the following interesting interpretation of the Milne
universe \cite{65}, originally due to \mbox{Milne \cite{61,62,63,64}.}

In the Minkowski space there is a point-like explosive event, the Big Bang,
resulting in infinitely many debris of point particles (fundamental observers)
shot out at all speeds up to the speed of light in all directions. It is
possible to arrange a special kind of velocity distribution of fundamental
observers for which the cosmological principle will remain true: all
fundamental observers will see the identical velocity distributions. Indeed,
let us show that this is possible \cite{62}. Let $f(u)$ be such a distribution.
Because of the assumed isotropy of space, this distribution can depend only
on the magnitude $u$ of the velocity $\vec{u}$, not on its direction. The
distribution function $f(u)$ does not depend on time because after the Big
Bang no new fundamental particles are created or destroyed and each of them
moves with a constant velocity. The cosmological principle demands
\begin{equation}
f(u^\prime)\;du_x^\prime du_y^\prime du_z^\prime=f(u)\;du_x du_y du_z
\label{eq60}
\end{equation}
where (according to our conventions, the light velocity $c=1$)
\begin{equation}
\vec{u}^{\,\prime}=\vec{u}\oplus \vec{v}=\frac{1}{1+\vec{u}\cdot \vec{v}}
\left [ \frac{\vec{u}}{\gamma_v}+\left(1+\frac{\gamma_v}{1+\gamma_v}\vec{u}
\cdot \vec{v}\right )\vec{v}\right ]
\label{eq61}
\end{equation}
is given by the relativistic velocity addition formula (see, for example,
\cite{66}). The Jacobian of this transformation is (to avoid involved
calculations by hand, REDUCE Computer Algebra System \cite{67} can be used
to get this result)
$$\frac{\partial(u_x^\prime, u_y^\prime, u_z^\prime)}
{\partial(u_x, u_y, u_z)}=\frac{1}{\gamma_v^4(1+\vec{u}\cdot \vec{v})^4}=
\left(\frac{\gamma_{u}}{\gamma_{u^\prime}}\right )^4$$
where at the last step we have used the so called gamma identity \cite{68,69}
$$\gamma_{\vec{u}\oplus\vec{v}}=\gamma_u\gamma_v (1+\vec{u}\cdot \vec{v})$$
Therefore, Eq. (\ref{eq60}) takes the form
$$\frac{f(u^\prime)}{\gamma_{u^\prime}^4}=
\frac{f(u)}{\gamma_u^4}$$
and we see that (we have restored the light velocity $c$ in the last formula
for a moment)
\begin{equation}
f(u)=B\;\gamma_u^4=\frac{B}{c^3\left(1-u^2/c^2\right)^2}
\label{eq62}
\end{equation}
with some positive constant $B$. This velocity distribution function has
several remarkable properties \cite{62}. The total number of particles
$N=\int f(u)\;du_x du_y du_z$ is infinite and the velocity-centroid (the
mean velocity) $<\vec{u}>=\frac{1}{N}\int \vec{u}\;f(u)\;du_x du_y du_z$
cannot be defined. The density of particles (in velocity-space) increases
and tends to infinity as the velocity $u$ approaches the light velocity.
Each fundamental observer can be regarded as the stationary center of an
expanding ball which looks one and the same for every choice of the
fundamental observer. Not every explosion readily produces such kind of
distribution. In fact,  ``it must have been an incredibly delicately tuned
Big Bang to achieve this!'' \cite{65}.

The Milne distribution Eq. (\ref{eq62}) is not of statistical  but of
hydrodynamic nature: the velocity $\vec{u}$ of every fundamental particle
at time $T$ is uniquely determined by its position through the 
\mbox{Hubble-like law}
$$ u_x=\frac{X}{T},\;\;u_y=\frac{Y}{T},\;\;u_y=\frac{Z}{T}$$

This fact allows to rewrite Eq. (\ref{eq62}) as a spatiotemporal 
distribution
$$dN=f(u)\;du_x du_y du_z=\frac{B}{\left(1-R^2/T^2\right)^2}\;\frac{dX dY dZ}
{T^3}=n(R,T)\;dX dY dZ$$
with the particle density function
\begin{equation}
n(R,T)=\frac{B\;T}{(T^2-R^2)^2}
\label{eq63}
\end{equation}

The density tends to infinity at the unattainable boundary $R=T$ of the
expanding ball which moves away at the light velocity.

The density Eq. (\ref{eq63}) is inhomogeneous and therefore the Copernican
principle is not evident in the Minkowski coordinates $R$ and $T$. Of course,
indirectly it is implemented through the equivalence of fundamental
observers (cosmological principle). However, for the Milne
universe there exists a natural foliation of space-time under which the
spatial distribution of matter becomes homogeneous and thus the Copernican
principle becomes obvious.

The symmetry group of special relativity (Minkowski space-time) is the
ten parameter Poincar\'{e} group of inhomogeneous (including space-time
translations) Lorentz transformations. The symmetry group of the Milne
universe is a six parameter subgroup of the Poincar\'{e} group of homogeneous
Lorentz transformations which leave invariant a special event---the Big
Bang. As a result, there is a preferred class of inertial observers,
namely those whose world-lines pass through the Big Bang event. The Milne
universe is a  description of the Minkowski space-time from the point of view
of members of this preferred class \cite{70}. The invariant proper time
after the Big Bang
\begin{equation}
t=\sqrt{T^2-X^2-Y^2-Z^2}=\sqrt{T^2-R^2}
\label{eq64}
\end{equation}
measured by the clocks of this fundamental observers, can be used for
a natural definition of simultaneity $t=\mathrm{const}$ in the Milne universe,
alternative to the usual Einsteinian simultaneity $T=\mathrm{const}$
(interestingly, Einstein's definition of simultaneity by means of the light
signal exchanges coincides, in principle, with the St. Augustine's criterion
of simultaneity given by him 1500 years earlier \cite{71}).

This cosmic time $t$ leads to the slicing (of a part) of the Minkowski
space-time into space-like hypersurfaces Eq. (\ref{eq64}) (two-sheeted
hyperboloids $T^2-X^2-Y^2-Z^2=t^2$) which Milne calls the public space to
distinguish it from the private space of fundamental observers defined
through the $T=\mathrm{const}$ slicing, $T$ being the usual Minkowski time.
Note that immediately after the Big Bang (for every $t>0$) the public space
is infinite, while the private space for every $T>0$ is a finite sphere of
radius $R=T$, increasing in size with light velocity as $T$ goes on. Thus,
the Milne model provides an excellent demonstration of a seemingly paradoxical
situation that a point-like Big Bang can give birth to a universe with
infinite spatial extension \cite{65,72}. Interestingly, the Milne foliation
of Minkowski space-time is exactly the foliation which leads to the Dirac's
point-form of relativistic dynamics \cite{73,74}.

Let $\tilde{n}(\psi,t)$ be the density of particles in public space so that
the number of particles in the volume element
$dV=\sqrt{|\tilde{\bf{g}}|}\;d\psi\;d\theta\;d\phi$ at cosmic time $t$ is
$dN=\tilde{n}(\psi,t)dV$. For the metric
Eq. (\ref{eq57}), the spatial components of the metric tensor are
$$\tilde{\bf{g}}_{\psi\psi}=t^2,\;\;\tilde{\bf{g}}_{\theta\theta}=
t^2\sinh^2{\psi},\;\;\tilde{\bf{g}}_{\phi\phi}=t^2\sinh^2{\psi}
\sin^2{\theta}$$

Therefore, the square root of the absolute value of the determinant of the
metric tensor equals to $\sqrt{|\tilde{\bf{g}}|}=t^3\sinh^2{\psi}\sin{\theta}$
and
\begin{equation}
dN=\tilde{n}(\psi,t)\;t^3\sinh^2{\psi}\sin{\theta}\;d\psi\;d\theta\;d\phi
\label{eq65}
\end{equation}

On the other hand, let us calculate the same number of particles in the
private space. For fixed cosmic time $t$, the radial coordinates lay
in-between $R=t\sinh{\psi}$ and $R+dR$ for the selected bunch of particles
in the private space, where $dR=t\cosh{\psi}\;d\psi$. But, according
to Eq. (\ref{eq58}), this particles have different Minkowski time 
coordinates
$T=t\cosh{\psi}$. Namely, the particles with the radial coordinate $R+dR$
have the Milne coordinate $\psi+d\psi$ and hence the $T$-time coordinate
$T+dT$ with $dT=t\sinh{\psi}\;d\psi$. However, in private space the
fundamental particles move radially outward with velocity $V=\tanh{\psi}$.
Therefore, at time $T$ the bunch of particles have a reduced radial spread
$$\widetilde{dR}=dR-VdT=t\cosh{\psi}\;d\psi-t\tanh{\psi}\;\sinh{\psi}\;d\psi=
\frac{t}{\cosh{\psi}}\;d\psi$$
and hence occupy $\widetilde{dV}=R^2\sin{\theta}\;\tilde{dR}\;d\theta\;d\phi$
volume element in the private space. The corresponding number of particles is
\begin{equation}
dN=n(R,T)\;\widetilde{dV}=n(R,T)\;\frac{t^3}{\cosh{\psi}}\;\sinh^2{\psi}\;
\sin{\theta}\;d\psi\;d\theta\; d\phi
\label{eq66}
\end{equation}

From Eq.s (\ref{eq65}) and  (\ref{eq66}), we get
\begin{equation}
\tilde{n}(\psi,t)=\frac{n(R,T)}{\cosh{\psi}}=\frac{B}{t^3}
\label{eq67}
\end{equation}
where the last step follows from equations Eq.s (\ref{eq58}), 
(\ref{eq63}) and
(\ref{eq64}). As we see, the density $\tilde{n}$ depends only on cosmic time
$t$ and not on the position in the public space---the distribution of
fundamental particles in the public space is not only isotropic but also
homogeneous and the Copernican  principle becomes explicit.

In more formal way, the same result can be obtained as follows \cite{70}.
The density-flow four-vector
\begin{equation}
J^\mu=(J^T,J^X,J^Y,J^Z)=
(n,nu_x,nu_y,nu_z)=\left (n,n\frac{X}{T},n\frac{Y}{T},n\frac{Z}{T}
\right)
\label{eq68}
\end{equation}
where the density $n$ is given by Eq. (\ref{eq63}), transforms under the 
change of coordinates

\begin{eqnarray} &&
T=t\cosh{\psi},\;\;X=t\sinh{\psi}\;\sin{\theta}\;\cos{\phi}\nonumber \\ &&
Y=t\sinh{\psi}\;\sin{\theta}\;\sin{\phi},\;\;
Z=t\sinh{\psi}\;\cos{\theta}\nonumber
\end{eqnarray}
as follows
\begin{eqnarray} &&
j^t=\frac{\partial{t}}{\partial{T}}\;J^T+\frac{\partial{t}}{\partial{X}}\;J^X
+\frac{\partial{t}}{\partial{Y}}\;J^Y+\frac{\partial{t}}{\partial{Z}}\;J^Z
\nonumber\\ &&
j^\psi=\frac{\partial{\psi}}{\partial{T}}\;J^T+\frac{\partial{\psi}}
{\partial{X}}\;J^X+\frac{\partial{\psi}}{\partial{Y}}\;J^Y+
\frac{\partial{\psi}}{\partial{Z}}\;J^Z
\nonumber \\ &&
j^\theta=\frac{\partial{\theta}}{\partial{T}}\;J^T+\frac{\partial{\theta}}
{\partial{X}}\;J^X+\frac{\partial{\theta}}{\partial{Y}}\;J^Y+
\frac{\partial{\theta}}{\partial{Z}}\;J^Z
\nonumber \\ &&
j^\phi=\frac{\partial{\phi}}{\partial{T}}\;J^T+\frac{\partial{\phi}}
{\partial{X}}\;J^X+\frac{\partial{\phi}}{\partial{Y}}\;J^Y+
\frac{\partial{\phi}}{\partial{Z}}\;J^Z
\label{eq69}
\end{eqnarray}

Differentiating $T^2-R^2=t^2$, we get
$$\frac{\partial{t}}{\partial{T}}=\frac{T}{t},\;\;
\frac{\partial{t}}{\partial{X}}=-\frac{X}{t},\;\;
\frac{\partial{t}}{\partial{Y}}=-\frac{Y}{t},\;\;
\frac{\partial{t}}{\partial{Z}}=-\frac{Z}{t}$$
which gives
$$j^t=\frac{T}{t}n-\frac{R^2}{tT}n=n\frac{t}{T}=\frac{B}{t^3}$$

Other portion of the partial derivatives can be obtained by
differentiating $R/T=\tanh{\psi}$. The \mbox{results are}
$$\frac{\partial{\psi}}{\partial{T}}=-\frac{R}{t^2},\;\;
\frac{\partial{\psi}}{\partial{X}}=\frac{XT}{Rt^2},\;\;
\frac{\partial{\psi}}{\partial{Y}}=\frac{YT}{Rt^2},\;\;
\frac{\partial{\psi}}{\partial{Z}}=\frac{ZT}{Rt^2}$$
and correspondingly we get from Eq. (\ref{eq69}) $j^\psi=0$.

Analogously, differentiating $\tan^2{\theta}=(X^2+Y^2)/Z^2$, we get
\begin{eqnarray} &&
\frac{\partial{\theta}}{\partial{T}}=0,\;\;
\frac{\partial{\theta}}{\partial{X}}=\frac{XZ}{R^2\sqrt{X^2+Y^2}}
\nonumber \\ &&
\frac{\partial{\theta}}{\partial{Y}}=\frac{YZ}{R^2\sqrt{X^2+Y^2}},\;\;
\frac{\partial{\theta}}{\partial{Z}}=-\frac{\sqrt{X^2+Y^2}}{R^2}
\end{eqnarray}
while differentiating $\tan{\phi}=Y/X$, we get
$$\frac{\partial{\phi}}{\partial{T}}=0,\;\;
\frac{\partial{\phi}}{\partial{X}}=-\frac{Y}{X^2+Y^2},\;\;
\frac{\partial{\phi}}{\partial{Y}}=\frac{X}{X^2+Y^2},\;\;
\frac{\partial{\theta}}{\partial{Z}}=0$$

Substituting these partial derivatives in Eq. (\ref{eq69}), we find that
$j^\theta=0$ and $j^\phi=0$. Therefore, in the Milne universe the density-flow
four-vector has the form
\begin{equation}
j^\mu=(J^t,J^\psi,J^\theta,J^\phi)=\left (\frac{B}{t^3},0,0,0 \right)
\label{eq70}
\end{equation}

We see that matter (fundamental particles) is at rest and uniformly
distributed in the Milne Universe. As was already mentioned, the Milne
distribution Eq. (\ref{eq62}) does not allow to determine the mean 
velocity of
the whole system of fundamental particles and therefore the global standard
of rest. However, using the mean motion of matter in the limited portion of
space-time, we can define a local standard of rest in the neighborhood of
a given point by defining rest and velocity relative to this mean motion.
Introduction of the cosmic time and public space (co-moving coordinates) is
just a realization of this strategy \cite{70}.

In the following, for simplicity, we will use a two-dimensional space-time
to illustrate our arguments. The two-dimensional Milne metric has the form
($t\ge 0,\;-\infty<\psi<\infty$)
\begin{equation}
ds^2=dt^2-t^2d\psi^2
\label{eq71}
\end{equation}

How can imaginary intelligent observers in the Milne universe discover
that their space-time is just a part of the more vast Minkowski space-time?
The coordinate transformation Eq. (\ref{eq58}) is not simple enough to be
trivially guessed. Instead we can envisage the following line of thoughts
(adapted \mbox{from \cite{74}).}

Null geodesics (light rays) can be used to construct a natural coordinate
grid in the Milne universe. Coordinates with respect to this grid (null
coordinates) are expected to convey and reveal the intrinsic geometry of
the Milne space-time. According to Eq. (\ref{eq71}), null geodesics are
determined by \mbox{the condition}
$$\left(\frac{d\psi}{dt}\right)^2=\frac{1}{t^2}$$

Integrating, we get that along null geodesics
$$\psi=\pm\ln{(gt)}$$
where $g>0$ is an integration constant such that $gt$ is dimensionless.
Accordingly, we introduce \mbox{null coordinates}
\begin{equation}
u=\psi-\ln{(gt)},\;\;\;v=\psi+\ln{(gt)}
\label{eq72}
\end{equation}

The inverse transformation is
$$\psi=\frac{1}{2}(u+v),\;\;\;t=\frac{1}{g}\;e^{(v-u)/2}$$
and after substituting it into Eq. (\ref{eq71}), we get
\begin{equation}
ds^2=-\frac{1}{g^2}\;e^{(v-u)}\;du\,dv
\label{eq73}
\end{equation}

The form of Eq. (\ref{eq73}) suggests to make further transformation
\begin{equation}
U=-e^{-u},\;\;\;V=e^v
\label{eq74}
\end{equation}
which brings the metric into the very simple form
\vspace{-24pt}
\begin{equation}
ds^2=-\frac{1}{g^2}\;dU\,dV
\label{eq75}
\end{equation}

Equation (\ref{eq74}) indicate that the Milne space-time corresponds to
the coordinate ranges $U<0$ and $V>0$. However the metric Eq. 
(\ref{eq75}) is
no longer singular and we can extend our space-time by allowing the whole
ranges  $-\infty<U<\infty$ and  $-\infty<V<\infty$. The identification
\begin{equation}
U=g(X-T),\;\;\;V=g(X+T)
\label{eq76}
\end{equation}
puts the metric in the canonical Minkowskian form $ds^2=dT^2-dX^2$ and thus
shows that our extended space-time is just the Minkowski space-time.
The coordinate transformation, inverse to \mbox{Eq. (\ref{eq76}), is}
\begin{equation}
X=\frac{1}{2g}(U+V),\;\;\;T=\frac{1}{2g}(V-U)
\label{eq77}
\end{equation}
which implies the following relations between the Minkowski coordinates
$(T,X)$ and the Milne coordinates $(t,\psi)$ (compare with Eq. 
(\ref{eq58}))
\begin{equation}
T=t\cosh{\psi},\;\;\;X=t\sinh{\psi}
\label{eq78}
\end{equation}

If we choose instead $U=e^{-u},\;\;\;V=-e^v$, we get the contacting
Milne universe with
\begin{equation}
T=-t\cosh{\psi},\;\;\;X=-t\sinh{\psi}
\label{eq79}
\end{equation}
which correspond to the $U>0,\,V<0$ patch of the Minkowski space-time. What
about $U>0,\,V>0$ and $U<0,\,V<0$ patches of the extended space-time?
Let us take
\begin{equation}
U=e^{-\tilde{u}},\;\;\;V=e^{\tilde{v}},\;\;\mathrm{with}\;\;
\tilde{u}=\alpha-\ln{\beta},\;\;\;\tilde{v}=\alpha+\ln{\beta}
\label{eq80}
\end{equation}

Substituting into Eq. (\ref{eq75}), we get
$$ds^2=\frac{\beta^2}{g^2}\left (d\alpha^2-\frac{d\beta^2}{\beta^2}\right )$$

Therefore, $\alpha$ is timelike and $\beta$ is spacelike. Thus, we identify
$\alpha=gt_R=\psi,\,\beta=gx_R$ which brings the metric into the form
\begin{equation}
ds^2=g^2x_R^2dt_R^2-dx_R^2=x_R^2d\psi^2-dx_R^2
\label{eq81}
\end{equation}

Such a metric is known as Rindler metric \cite{6}. According to Eq. 
(\ref{eq77}), the Minkowski coordinates $(T,X)$ and Rindler coordinates 
$(t_R,\,x_R)$ are related as follows
\begin{equation}
T=x_R\;\sinh{(gt_R)},\;\;\;X=x_R\;\cosh{(gt_R)}
\label{eq82}
\end{equation}

It is clear from Eq. (\ref{eq80}) that we must have $x_R>0$. Therefore the
Rindler space-time corresponds to the $X>0,\,|T|<X$ portion of the Minkowski
space-time (right Rindler wedge). The choice  $U=-e^{-\tilde{u}},\;\;\;
V=-e^{\tilde{v}}$ leads to the left Rindler wedge with the same metric
Eq. (\ref{eq81}) \mbox{but with}

$$T=-x_R\;\sinh{(gt_R)},\;\;\;X=-x_R\;\cosh{(gt_R)}$$

\begin{figure}[ht]
 \begin{center}
    \includegraphics[scale=0.5]{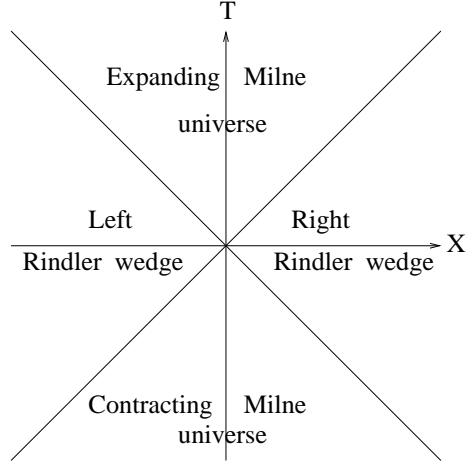}
  \end{center}
\caption{The patches of the Minkowski space-time covered by the Milne and
\mbox{Rindler metrics.}}
\label{fig3}
\end{figure}

Figure \ref{fig3} summarizes our findings. It is interesting to note that the
above discussion of the Milne (or Rindler) space-time extension closely
resembles the celebrated Kruskal extension of the Schwarzschild space-time
\cite{6,75,76}. More precisely, the expanding Milne universe corresponds to
the black hole region of the Kruskal diagram, the contracting Milne universe
corresponds to the white hole region, and the Rindler space-time corresponds
to outer Schwarzschild regions \cite{6}.

What is the physical meaning of the Rindler coordinates? Inspired by the Milne
interpretation outlined above, let us consider a set of Rindler fundamental
observers which are at rest in the Rindler space-time having constant $x_R$
spatial coordinates. The line of simultaneity for an arbitrary event
$P=(T,X)$ at the worldline of a Rindler observer is the set of events that
are simultaneous with $P$ in the inertial instantaneous rest frame of the
Rindler observer at $P$. As $x_R$ is fixed, this instantaneous rest frame
moves with the the Minkowski velocity (in the private space of the Big Bang
event)
\begin{equation}
v=\frac{dX}{dT}=\tanh{(gt_R)}=\frac{T}{X}
\label{eq83}
\end{equation}

Therefore, according to the Lorentz transformations, in the instantaneous rest
frame the event $P$ has the time coordinate $T^\prime=\gamma_v(T-vX)=0$,
what means that it is simultaneous with the Big Bang event. As both the
Rindler observer and the point on its worldline were arbitrary, we come to
an amusing conclusion that the Big Bang event is simultaneous with every event
on the worldline of every Rindler observer. In other words, the space axes of
the fundamental Rindler observers all pass through the Big Bang event.
In comparison, the defining characteristic property of the Milne observers is
that all their time axes pass through the Big Bang event. This interchange of
space and time makes a big difference: the Rindler observers cannot be
inertial because the space axes of the inertial observer are all parallel to
each other and thus cannot cross at the origin.

From Eq. (\ref{eq83}) the acceleration of the Rindler observer is
\begin{equation}
a=\frac{dv}{dT}=\frac{1}{X}\left (1-\frac{T^2}{X^2}\right )=\frac{x_R^2}{X^3}
\label{eq84}
\end{equation}

On the other hand, differentiating the velocity addition formula
$$v^\prime=\frac{v-V}{1-vV}$$
and using $T^\prime=\gamma(T-VX)$, we get
\begin{equation}
a^\prime=\frac{dv^\prime}{dT^\prime}=\frac{a}{\gamma^3(1-vV)^3}
\label{eq85}
\end{equation}

For the instantaneous rest frame $V=v$, so Eq.s (\ref{eq84}) and 
(\ref{eq85}) will give
\begin{equation}
a^\prime=\frac{\gamma^3x_R^2}{X^3}=\frac{1}{x_R}
\label{eq86}
\end{equation}
because
$$\gamma=(1-V^2)^{-1/2}=\left (1-\frac{T^2}{X^2}\right )^{-1/2}=
\frac{X}{x_R}$$

As we see from  Eq. (\ref{eq86}), the Rindler observers in the private 
space of the Big Bang event move with the constant proper acceleration which is
inversely proportional to their Rindler coordinate $x_R$ and approaches the
infinity for observers near the Big Bang, $x_R\to 0$.

Equation (\ref{eq83}) shows that at Rindler time $t_R$ all Rindler observers
have the same rapidity $\psi=gt_R$. In fact the rapidity is the most natural
(dimensionless) time coordinate in the Rindler space-time. The Rindler
metric expressed in terms of the rapidity  $\psi$ does not contain an
arbitrary constant $g$. Similarly, in the Milne space-time the rapidity is
the most natural (dimensionless) spatial coordinate.

What is the origin of the constant $g$? The fact that Rindler observers share
their lines of simultaneity enables us to define global synchronization in
the Rindler space-time: we can simply pick out an arbitrary Rindler observer
with the spatial coordinate $x_{R0}$ and declare its proper time to be the
coordinate time $t_R$. {But the proper time of the observer with the spatial
coordinate} $x_R$ is
$$\tau=\int\limits_0^T\frac{dT}{\gamma_v}=\int\limits_0^T\sqrt{1-\frac{T^2}
{X^2}}\;dT=x_R\int\limits_0^T\frac{dT}{\sqrt{x_R^2+T^2}}=x_R\;
\operatorname{arsinh}\frac{T}{x_R}$$

Comparing with Eq. (\ref{eq82}), we see that
\begin{equation}
\tau=(gx_R)\;t_R
\label{eq87}
\end{equation}

We have chosen to define synchronization $\tau=t_{R}$ for the selected
observer, and hence, $gx_{R0}=1$, so that $g$ is the proper acceleration
of this observer. Therefore, the freedom to use an arbitrary constant
$g=1/x_{R0}$ in description of the Rindler space-time is the freedom to
select any Rindler observer to define global synchronization and hence, the
coordinate time $t_R$. For Rindler observers other than the selected one,
the proper time will not coincide with the coordinate time but will be
proportional to it, as Eq. (\ref{eq87}) shows.

As we see, the Rindler space-time is a part of the Minkowski space-time
as viewed by a set of non-inertial observers whose constant proper
accelerations are adjusted in such a way that their lines of simultaneity
all pass through one event (Big Bang). Further details and a lucid discussion
of the physical meaning of Rindler  coordinates can be found in \cite{77}.

Note that in the four-dimensional space-time radially moving Rindler observers
define spherical Rindler coordinates $(t_R,r_R,\theta,\phi)$ as follows
\cite{78}
\begin{eqnarray} &&
T=r_R\sinh{(gt_R)},\;\;\; X=r_R\cosh{(gt_R)}\;\sin{\theta}\;\cos{\phi}
\nonumber \\ &&
Y=r_R\cosh{(gt_R)}\;\sin{\theta}\;\sin{\phi},\;\;\;
Z=r_R\cosh{(gt_R)}\;\cos{\theta}
\label{eq88}
\end{eqnarray}

They cover the whole external part of the Big Bang's light-cone (subdivision
into Rindler wedges arises only in the $(1+1)$-dimensional space-time) where
the induced metric is
$$ds^2=g^2r_R^2dt_R^2-dr_R^2-r_R^2\cosh^2{(gt_R)}(d\theta^2+\sin^2{\theta}\;
d\phi^2)$$

\section{St. Augustine's Objects}

\begin{figure}[htb]
 \begin{center}
    \includegraphics[scale=0.5]{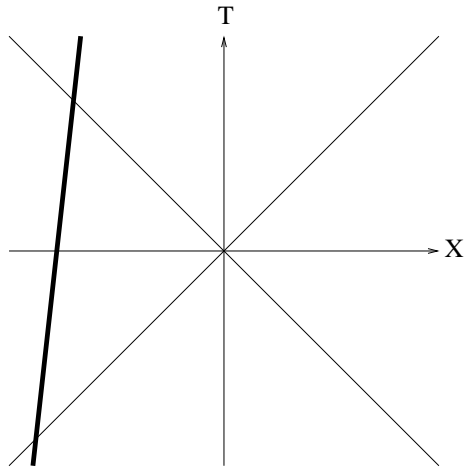}
  \end{center}
\caption{The world line of the St. Augustine's object drawn here as a thick
line.}
\label{fig4}
\end{figure}

Both Milne and Rindler fundamental observers may not suspect that their
space-times are just some parts of a bigger entity -- the Minkowski
space-time. However, we know that this is the case. A legitimate question
then is whether Milne and Rindler parts of the Minkowski space-time can
exchange material objects. As far as the space-time itself is concerned, the
only restriction is causality, because the Minkowski space-time is not
singular. Thus, naively, material objects from other parts of the
Minkowski space-time can penetrate expanding Milne universe. One timelike
world line of this kind is shown on Figure ~\ref{fig4}. We will call such
objects, maybe rather frivolously,  the St. Augustine's objects.

St. Augustine of Hippo was the most influential philosopher who pondered on
the nature of time. ``What then is time? If no one asks me, I know what it is.
If I wish to explain it to him who asks, I do not know'' -- ruminates he in his
{\it Confessions} \cite{79}. St. Augustine's version of the modern question
{\it what was going on before the big bang} sounds like this: ``What was God
doing before He made heaven and earth?''.  St. Augustine overcomes his own
temptation to laugh at the question and give a jesting answer: ``He was
preparing hell for those who pry into mysteries.'' Instead he comes to an
answer which strikingly resonates with the modern day ideas about space-time
and that the universe was created by quantum tunneling from a state with no
classical space time \cite{80,81}. Namely, in {\it City of God} \cite{82}
he comes to the conclusion that ``time does not exist without some movement
and transition'' and this is what distinguishes eternity and time. His main
inference then is that ``the world was made, not in time, but simultaneously
with time''. In modern words, time itself was created by the Big Bang and
the question what was before the Big Bang is meaningless, it is like asking
what is to the north to the North Pole \cite{83}.

However St. Augustine's brilliant insight that the time was created
simultaneously with the material world should be taken with a grain of salt,
because the use of the word ``create'' may imply that there was some concept
of time even at the epoch when time was still not existing  \cite{84,85}. The
Milne model provides a ready resolution of this dilemma.

As we know, there are two notions of time in the Milne model. The public time
is determined by an expanding substratum of fundamental particles. Space-like
hypersurfaces orthogonal to the world lines of these fundamental particles
define a public space. The public time is the proper time  along world lines
of the fundamental particles and this cosmic time in a sense labels the series
of the space-like hypersurfaces. Therefore, the public time is created by the
Big Bang and it is really meaningless to ask what happened before the public
time zero.

However, to any inertial fundamental particle in the Milne model we can attach
the usual Minkowski frame and thus define the private time of this observer.
If there are fundamental observers whose world lines begin not at the Big Bang
but in the preceding contracting phase, we can use the private time of such
observers supposing that they survived during the Big Bang, to make the
question of what happened before the Big Bang obviously meaningful.

The central question, therefore, is whether material objects can cross the
boundaries (the future light-cone of the Big Bang event) of the expanding
Milne universe. This question is not a trivial one, because the density of
the fundamental observers, as we have seen above,  becomes infinite at these
boundaries, as a result of the cosmological principle.

Milne himself was well aware of this problem. However, he considers the
question what is outside the sphere $R=T$ in the private space as meaningless
based on the following arguments \cite{62}.

Milne observers inside the sphere $R=T$ cannot observe any object outside,
because infinite density of fundamental particles on this sphere serves as
a curtain obscuring the vision and precludes any window into outer space in
any direction. Therefore, only objects which were overtaken by the expanding
frontier and appeared inside the sphere $R=T$ can be observed. For
inhabitants of the Milne universe the St. Augustine objects would suddenly
appear to observation in an act of creation. More precisely, from the point
of view of public time, such objects are just a part of the Big Bang initial
conditions.

However, Milne thinks that the infinite density of fundamental particles
on the expanding frontier can lead to the infinite number of collisions between
the St. Augustine object and the fundamental particles, thus, not allowing the
St. Augustine object to penetrate the boundary. In fact, not the infinite
density by itself is crucial in this argument, but a particular form of the
density function Eq. (\ref{eq63}), which implies that the total number of
particles in any thin layer including the boundary is infinite. Milne
concludes: ``such objects never come, and never can come, into interaction
with observable members of the given system. They may, mathematically
described, tend to be swept up by the expanding frontier, but they never
penetrate it. Penetration would be a logical self-inconsistency of
description of the whole system'' \cite{62}.

But can the ideal Milne universe actually be realized? The infinite number
of needed fundamental particles and the initial (in cosmic time) singularity
of infinite density, implied by Eq. (\ref{eq67}), speak against such 
a possibility. Another argument against the realizability of the ideal Milne 
universe can be obtained by considering quantum fields in the Milne universe.

In analogy with Eq. (\ref{eq77}), let us define the dimensionfull 
temporal and spatial $(t_M,x_M)$ \mbox{Milne coordinates:}
\begin{equation}
x_M=\frac{1}{2g} (u+v)=\frac{\psi}{g},\;\;\;t_M=\frac{1}{2g} (v-u)=
\frac{1}{g}\ln{(gt)}
\label{eq89}
\end{equation}

They are related to the Minkowski coordinates $(T,X)$ through:
\begin{equation}
gX=e^{gt_M}\sinh{(gx_M)},\;\;\;gT=e^{gt_M}\cosh{(gx_M)}
\label{eq90}
\end{equation}

The Milne metric Eq. (\ref{eq71}) in terms of these coordinates takes 
conformally flat form
\begin{equation}
ds^2=e^{2gt_M} (dt_M^2-dx_M^2)
\label{eq91}
\end{equation}

In contrast to the cosmic time $t$ whose range is $0\le t<\infty$, the range
of the conformal Milne time $t_M$ (as well as of the spatial coordinate $x_M$)
is the whole real axis from $-\infty$ to $\infty$.

Free scalar field $\Phi(t_M,x_M)$ with mass $m\ne 0$ in the Milne universe
satisfies the covariant Klein-Gordon equation
\begin{equation}
(\hat\square + m^2) \Phi=0
\label{eq92}
\end{equation}
where the covariant d'Alembertian (Laplace-Beltrami operator) has the form
\begin{equation}
\hat\square=\nabla_\mu\nabla^\mu=\nabla_\mu ({\bf{g}}^{\mu\nu}\nabla_\nu)
\label{eq93}
\end{equation}

For scalar $\Phi$ and vector $V^\mu={\bf{g}}^{\mu\nu}\partial_\nu \Phi$ fields
the covariant derivative $\nabla_\mu$ is defined as follows \cite{6}
$$\nabla^\mu \Phi=\partial^\mu \Phi,\;\;\;\nabla_\mu V^\nu=
\partial_\mu V^\nu+{\Gamma^\nu}_{\mu \sigma} V^\sigma$$
where the connection coefficients (Christoffel symbols) are
$${\Gamma^\nu}_{\mu \sigma}=\frac{1}{2}{\bf{g}}^{\nu\alpha}(\partial_\mu
{\bf{g}}_{\alpha\sigma}+ \partial_\sigma {\bf g}_{\mu \alpha}-\partial_\alpha
{\bf g}_{\sigma \mu})$$

Therefore,
\begin{equation}
\hat\square \Phi=\partial_\mu({\bf{g}}^{\mu\nu}\partial_\nu\Phi)+
{\Gamma^\mu}_{\mu \sigma}{\bf{g}}^{\sigma\nu}\partial_\nu\Phi=
{\bf{g}}^{\mu\nu}\partial_\mu\partial_\nu\Phi+[(\partial_\mu
{\bf{g}}^{\mu\nu})+{\Gamma^\mu}_{\mu \sigma}{\bf{g}}^{\sigma\nu}]
(\partial_\nu\Phi)
\label{eq94}
\end{equation}
and
$${\Gamma^\mu}_{\mu \sigma}=\frac{1}{2}\;{\bf{g}}^{\mu\alpha}[\partial_\mu
{\bf{g}}_{\alpha\sigma}+\partial_\sigma {\bf{g}}_{\mu\alpha}-\partial_\alpha
{\bf{g}}_{\sigma\mu}]
=\frac{1}{2}\;{\bf{g}}^{\mu\alpha}\partial_\sigma {\bf{g}}_{\mu\alpha}$$

For the metric Eq. (\ref{eq91}), the only non-zero components of the 
metric tensor are
$${\bf{g}}_{00}=e^{2gt_M},\;\;\; {\bf{g}}_{11}=-e^{2gt_M},\;\;\;
{\bf{g}}^{00}=e^{-2gt_M},\;\;\; {\bf{g}}^{11}=-e^{-2gt_M}$$

Then
$${\Gamma^\mu}_{\mu 0}=2g,\;\;\;{\Gamma^\mu}_{\mu 1}=0$$
and it can be easily checked that the second term in Eq. (\ref{eq94})
equals to zero. Finally,
$$\hat\square \Phi={\bf{g}}^{\mu\nu}\partial_\mu\partial_\nu\Phi=
e^{-2gt_M}\left(\frac{\partial^2}{\partial t_M^2}-
\frac{\partial^2}{\partial x_M^2}\right)\Phi$$
and the Klein-Gordon Eq. (\ref{eq92}) takes the form
\begin{equation}
\frac{\partial^2\Phi}{\partial t_M^2}-\frac{\partial^2\Phi}{\partial x_M^2}+
m^2e^{2gt_M}\Phi=0
\label{eq95}
\end{equation}

We can separate variables, taking $\Phi(t_M,x_M)=\phi(x_M)\psi(t_M)$, and get
$$\frac{\ddot{\psi}}{\psi}+m^2e^{2gt_M}=\frac{\phi^{\prime\prime}}{\phi}=
\lambda$$
where $\dot{\psi}=\frac{d\psi}{dt_M}$, $\phi^\prime=\frac{d\phi}{dx_M}$ and
$\lambda$ is some constant. Only if this constant is negative,
$\lambda=-k^2$, the solution $\phi(x_M)=e^{\pm ikx_M}$ does not grow
exponentially. Then the equation for $\psi(t_M)$ is
\begin{equation}
\ddot{\psi}+(k^2+m^2e^{2gt_M})\psi=0
\label{eq96}
\end{equation}

This equation is formally identical to the s-wave radial Schr\"{o}dinger
equation for an exponential potential with well known solution \cite{86}.
Namely, in terms of the cosmic time $t=\frac{1}{g}e^{gt_M}$,
\mbox{Eq. (\ref{eq96})} takes the form
\begin{equation}
\frac{d^2\psi}{dt^2}+\frac{1}{t}\,\frac{d\psi}{dt}+\left (\frac{k^2}{g^2t^2}+
m^2\right )\psi=0
\label{eq97}
\end{equation}
which is Bessel's equation (of pure imaginary order) with the general
solution \cite{86}
\begin{equation}
\psi=c_1J_{i\nu}(mt)+c_2J_{-i\nu}(mt)
\label{eq98}
\end{equation}
where $\nu=|k|/g$ and $c_1,c_2$ are arbitrary constants.

For small arguments \cite{87},
$$J_\nu(z)\approx\frac{1}{\Gamma(1+\nu)}\left(\frac{z}{2}\right )^\nu$$
and as $t_M\to -\infty$, so that $t\to 0$,
$$J_{\pm i\nu}(mt)\sim t^{\pm i\nu}=\left(\frac{1}{g}e^{gt_M}\right)^{\pm i
\frac{|k|}{g}}\sim e^{\pm i|k|t_M}$$

Therefore, to get the positive-frequency mode (which behaves as
$e^{- i|k|t_M}$) in the asymptotic past (in Milne conformal time $t_M$), we
must take $c_1=0$. The second constant, $c_2$, can be fixed by normalizing
the mode using the covariant norm \cite{88,89}
\begin{equation}
(\Psi_1,\Psi_2)=-i\int\limits_\Sigma \sqrt{-{\bf{g}}}\, d\Sigma^\mu \,(\Psi_1
\partial_\mu \Psi_2^*-\Psi_2^*\partial_\mu\Psi_1)
\label{eq99}
\end{equation}
where ${\bf{g}}=\mathrm{det}({\bf{g}}_{\mu\nu})$ is the determinant of
the metric tensor, $\Sigma$ is a spacelike hypersurface with $n^\mu$ being the
unit timelike vector normal to it,  and
$\sqrt{-{\bf{g}}} \, d\Sigma$ is the invariant (proper) volume element in
this hypersurface. Using the Gauss' theorem for curved manifolds \cite{90},
$$\int\limits_V(\nabla_\mu j^\mu)\, \sqrt{-{\bf{g}}}\,dV=
\oint\limits_{\partial V}\sqrt{-{\bf{g}}}\,j^\mu d\Sigma_\mu$$
it can be proved \cite{88,89} that the scalar product is independent of the
choice of spacelike surface $\Sigma$  provided $\Psi_1$ and $\Psi_2$
are solutions of the Klein-Gordon Eq. (\ref{eq92}) that vanish
sufficiently quickly at spatial infinity. Indeed, let $V$ be the four-volume
whose boundary $\partial V$ consists of non-intersecting spacelike
hypersurfaces $\Sigma_1$, $\Sigma_2$ and, possibly, by timelike boundaries at
spatial infinity on which \mbox{$\Psi_1=\Psi_2=0$.} Then we have
$$(\Psi_1,\Psi_2)_{\Sigma_1}-(\Psi_1,\Psi_2)_{\Sigma_2}=
i\oint\limits_{\partial V}\sqrt{-{\bf{g}}}\,j^\mu d\Sigma_\mu=
i\int\limits_V(\nabla^\mu j_\mu)\, \sqrt{-{\bf{g}}}\,dV$$
where $j_\mu=\Psi_2^*\partial_\mu\Psi_1-\Psi_1\partial_\mu\Psi_2^*=
\Psi_2^*\nabla_\mu\Psi_1-\Psi_1\nabla_\mu\Psi_2^*$. But as $\Psi_1$ and
$\Psi_2$ are solutions of the Klein-Gordon Eq. (\ref{eq92}), then:
$$\nabla^\mu j_\mu=\Psi_2^*\hat\square\Psi_1-\Psi_1\hat\square\Psi_2^*=0$$

Therefore,
$$(\Psi_1,\Psi_2)_{\Sigma_1}=(\Psi_1,\Psi_2)_{\Sigma_2}$$

Let us specify the scalar product Eq. (\ref{eq99}) for our case of 
two-dimensional Milne space-time with the metric Eq. (\ref{eq91}). 
The convenient choice of
$\Sigma$ is the spacelike hyperbola defined by the condition
$t_M=\mathrm{const}$. Then the proper line element in $\Sigma$ is
$\sqrt{-{\bf{g}}} \,d\Sigma=t\,d\psi=e^{gt_M}dx_M$,
and the unit vector (normalized according to the metric Eq. 
(\ref{eq91})) is
$n^\mu=(e^{-gt_M},0)$. Therefore, the scalar product takes the form
\begin{equation}
(\Psi_1,\Psi_2)=i\int\limits_{-\infty}^\infty dx_M \,
\left(\Psi_2^*\frac{\partial \Psi_1}{\partial t_M}-
\Psi_1\frac{\partial \Psi_2^*}{\partial t_M}\right)
\label{eq100}
\end{equation}

Now we can determine the constant $c_2$. In fact, using $J^*_{i\nu}(mt)=
J_{-i\nu}(mt)$, the Wronskian \mbox{relation \cite{87}}
$$W\{J_\nu (z),J_{-\nu} (z)\}=J_\nu\;\frac{d J_{-\nu}}{dz}-J_{-\nu}\;
\frac{d J_\nu}{dz}=-\frac{2}{\pi z}\,\sin{(\nu \pi)}$$
and the normalization (for the positive-frequency modes)

$$(\Psi_k,\Psi_{k^\prime})=\delta(k-k^\prime)$$
we find that the normalized mode, which behaves as of pure positive frequency
in the asymptotic past, has the form
\begin{equation}
\Psi_{k,in}^{(+)}(t_M,x_M)=\frac{1}{2\sqrt{g\,\sinh{(\nu\pi)}}}\,e^{ikx_M}
J_{-i\nu}\left(\frac{m}{g}\,e^{gt_M}\right )
\label{eq101}
\end{equation}

In the asymptotic future, $t_M\to\infty$, the cosmic time $t$ also tends to
infinity and we should use the following asymptotic behavior \cite{87}:
when $|z|\to\infty,\,|\arg{z}|<\pi$, then
\begin{equation}
J_\nu(z)\approx\sqrt{\frac{2}{\pi z}}\,\cos{\left (z-\frac{1}{2}\nu\pi-
\frac{1}{4}\pi\right)}
\label{eq102}
\end{equation}

Using Eq. (\ref{eq102}), it can be verified that the combinations  
Eq. (\ref{eq98})
that correspond to the positive and negative frequency behavior
($\sim e^{\mp i mt}$) in asymptotic future are, respectively, the Hankel
functions $H^{(2)}_{i\nu}(mt)$ and $H^{(1)}_{i\nu}(mt)$, defined by
\begin{equation}
H^{(1)}_{i\nu}(z)=\frac{e^{\nu\pi}J_{i\nu}(z)-J_{-i\nu}(z)}{\sinh{(\nu\pi)}},
\;\;\;
H^{(2)}_{i\nu}(z)=\frac{J_{-i\nu}(z)-e^{-\nu\pi}J_{i\nu}(z)}{\sinh{(\nu\pi)}}
\label{eq103}
\end{equation}

{Using the Wronskian relation} \cite{87}
$$W\{H^{(1)}_\nu (z),H^{(2)}_{\nu} (z)\}=-\frac{4i}{\pi z}$$
we find that the normalized modes, corresponding to the  positive and
negative frequencies in the asymptotic future, have the form
\begin{eqnarray} &&
\Psi_{k,out}^{(+)}(t_M,x_M)=\frac{1}{\sqrt{8g}}\,e^{ikx_M}\,e^{\nu\pi/2}\,
H^{(2)}_{i\nu}\left(\frac{m}{g}\,e^{gt_M}\right )\nonumber\\ &&
\Psi_{k,out}^{(-)}(t_M,x_M)=\frac{1}{\sqrt{8g}}\,e^{ikx_M}\,e^{-\nu\pi/2}\,
H^{(1)}_{i\nu}\left(\frac{m}{g}\,e^{gt_M}\right )
\label{eq104}
\end{eqnarray}

Note that the negative frequency mode has the negative norm:
$$(\Psi^{(-)}_k,\Psi^{(-)}_{k^\prime})=-\delta(k-k^\prime).$$
Equations (\ref{eq101}), (\ref{eq103}) and (\ref{eq104}) imply the relation
\begin{equation}
\Psi_{k,in}^{(+)}=\frac{1}{\sqrt{2\sinh{\nu\pi}}}\left [
e^{\nu\pi/2}\Psi_{k,out}^{(+)}+e^{-\nu\pi/2}\Psi_{k,out}^{(-)}\right ]
\label{eq105}
\end{equation}
which shows that the pure positive-frequency mode $\Psi_{k,in}^{(+)}$ in the
asymptotic past evolves into a superposition of positive- and
negative-frequency modes $\Psi_{k,out}^{(+)}$ and $\Psi_{k,out}^{(-)}$
of the asymptotic future \cite{91}. The standard interpretation \cite{88}
of this fact is that the time-dependent background metric leads to a pair
production with the averaged number density of produced pairs in the
$k$-mode
$$n(k)=\left |\int\limits_{-\infty}^\infty (\Psi_{k,out}^{(-)},
\Psi_{k^\prime,in}^{(+)})\,dk^\prime\right |^2=\left |\frac{-e^{-\nu\pi/2}}
{\sqrt{2\sinh{\nu\pi}}}\right |^2=\frac{1}{e^{\frac{2\pi}{g}|k|}-1}$$

The problem, however, is that, if we consider the Milne universe as just
a re-parametriza\-tion of the part of Minkowski space-time, the time dependence
of the background metric is just a coordinate artifact, not related to the
real presence of a variable gravitational field, and, hence, no particle
production is \mbox{expected \cite{91}}. The crux of the problem can be traced 
back to the definition of the initial vacuum \mbox{state \cite{92}.}

Suppose we have a nonzero rate $R$ of particle production due to
time-variable background metric. To measure precisely the number of particles
in some volume, the measurement duration $\Delta t$ must be sufficiently small,
namely such that $|R|\Delta t\ll 1$. However, the smaller is the measurement
time $\Delta t$, the greater is the uncertainty in energy $\Delta E\sim
1/\Delta t$, and due to the possibility of temporary creation of virtual
particle pairs, the number of particles will become uncertain by the amount
$\Delta E/m\sim 1/(m\Delta t)$. Therefore, over a time interval $\Delta t$,
{the total uncertainty in the number of particles is} \cite{88,93}
$$\Delta N=\frac{1}{m\Delta t}+|R|\,\Delta t$$

{The optimum is reached at}
$$\Delta t =\frac{1}{\sqrt{m|R|}}$$
which gives the following minimum inherent uncertainty of the number of
particles:
\begin{equation}
\Delta N_m=2\sqrt{\frac{|R|}{m}}
\label{eq106}
\end{equation}

This simple computation indicates that in an adiabatic region, where the
Hubble parameter $H=\dot{a}/a$ is small (the scale factor $a(t)$ varies
slowly and hence $|R|$ is expected to be small), the concept of particles,
and hence of the vacuum state, is a well
defined concept. For the Milne universe $H=1/t$. Therefore, the vacuum state
is well defined in remote future. However, the region of the Milne space-time
near the initial singularity is not an adiabatic region and we do not have
a natural well defined vacuum state there. We can use (as we have, in fact,
done above) the conformal time $t_M$ and quantum states that contain
only positive frequencies with respect to this conformal time, when
$t_M\to -\infty$, to define the initial vacuum state in this region. Such
conformal vacuum state is, of course, not adiabatic and its use
in the role of the initial vacuum state may seem as a somewhat arbitrary
choice \cite{92}. Nevertheless, this choice is justified in the path-integral
formulation, where it corresponds to the natural restriction of the
paths summed to those that lie inside the future light-cone of the initial
singularity (the Big Bang event) \cite{94}.

However, this restriction is natural only when the initial singularity is real
and not a coordinate artifact, as in the case of the naive interpretation of
the Milne model.

When $x>0$, we have the following integral representations of the Hankel
functions \cite{95}:
\begin{equation}
H_\nu^{(1)}(x)=\frac{e^{-\frac{1}{2}i\nu\pi}}{i\pi}\int\limits_{-\infty}
^\infty e^{ix\cosh{u}-\nu u}du,\;
H_\nu^{(2)}(x)=-\frac{e^{\frac{1}{2}i\nu\pi}}{i\pi}\int\limits_{-\infty}
^\infty e^{-ix\cosh{u}-\nu u}du
\label{eq107}
\end{equation}

Using these integral representations in Eq. (\ref{eq104}), we get
\begin{eqnarray} &&
\Psi_{k,out}^{(+)}=\frac{i}{\pi\sqrt{8g}}\int\limits_{-\infty}
^\infty e^{-imt\cosh{u}-i\frac{k}{g}(u-gx_M)}du\nonumber \\ &&
\Psi_{k,out}^{(-)}=-\frac{i}{\pi\sqrt{8g}}\int\limits_{-\infty}
^\infty e^{imt\cosh{u}-i\frac{k}{g}(u-gx_M)}du
\label{eq108}
\end{eqnarray}

After changing the integration variable $u$ in the above integrals
respectively to $\tilde\psi=\pm(gx_m-u)=\pm (\psi-u)$, Eq.s (\ref{eq108})
take the form
\begin{eqnarray} &&
\Psi_{k,out}^{(+)}=\frac{i}{\pi\sqrt{8g}}\int\limits_{-\infty}^\infty
e^{-i(ET-PX)+i\frac{k}{g}\tilde\psi}d\tilde\psi\nonumber \\ &&
\Psi_{k,out}^{(-)}=-\frac{i}{\pi\sqrt{8g}}\int\limits_{-\infty}^\infty
e^{i(ET+PX)-i\frac{k}{g}\tilde\psi}d\tilde\psi
\label{eq109}
\end{eqnarray}
where $E=m\cosh{\tilde\psi}=\sqrt{m^2+P^2}$ and $P=m\sinh{\tilde\psi}$.
As we see, $\Psi_{k,out}^{(+)}$ is a superposition of positive frequency
Minkowski plane waves with different rapidities $\tilde\psi$, while
$\Psi_{k,out}^{(-)}$ is a superposition of negative frequency
Minkowski plane waves \cite{88,96}. Of course, instead of the rapidity
$\tilde\psi$, we can use the momentum $P=m\sinh{\tilde\psi}$ as an
integration variable \cite{92}:
\begin{eqnarray}
\Psi_{k,out}^{(+)}&=&\frac{i}{\pi\sqrt{8g}}\int\limits_{-\infty}^\infty
e^{-i(\sqrt{m^2+P^2}\,T-PX)+i\frac{k}{g}\operatorname{arsinh}{\frac{P}{m}}}
\;\frac{dP}{\sqrt{m^2+P^2}}\nonumber \\
\Psi_{k,out}^{(-)}&=&-\frac{i}{\pi\sqrt{8g}}\int\limits_{-\infty}^\infty
e^{i(\sqrt{m^2+P^2}\,T+PX)-i\frac{k}{g}\operatorname{arsinh}{\frac{P}{m}}}
\;\frac{dP}{\sqrt{m^2+P^2}}
\label{eq110}
\end{eqnarray}

The integral representations Eq. (\ref{eq109}), or Eq. (\ref{eq110}) 
define a natural analytic continuations of $\Psi_{k,out}^{(+)}$ and 
$\Psi_{k,out}^{(-)}$ on
the whole Minkowski space-time \cite{92,96} and, of course, these analytic
continuations pick up a privileged vacuum state corresponding to no particle
production. If we consider the Milne universe as a part of Minkowski
space-time, this choice of the vacuum state is the most natural.
However, from the point of view of the inside observers of the Milne
universe, it may appear rather contrived and will require subtle arguments
to justify it from the inside perspective \cite{92}.

An other possibility is given by considering the Milne universe as a limiting
case of some nonsingular space-time whose metric differs from the
Milne-Rindler metric only in the narrow transition region around the Big-Bang
light-cone \cite{97}. In this case the particle production can be related to
the nonzero time-variable curvature of the space-time (a real time dependent
gravitational field) in the transition region. One can imagine that this
effect of particle production can take place even in the limit of zero width
of the transition region, however such a situation will require a physically
unrealistic sources of infinite power on the Milne universe singular
boundary \cite{97}. Of course, this picture is in accord with the infinite
density of the Milne observers at the boundary, but these infinities make
the construction of the ideal Milne universe with impenetrable boundaries
(for example, in hydrodynamical laboratory experiments to mimic an arbitrary
FRW space-time by relativistic acoustic geometry \cite{98}) impossible.
Therefore, any realistic approximation to the ideal Milne universe is expected
to have boundaries penetrable for St. Augustine's objects.

Now we assume that the same is true to our real universe: the initial
singularity is only an unrealistic idealization and in realistic settings
it is traversable for St. Augustine's objects. As we expect that the initial
density at the Big Bang was nevertheless very high, we assume that only black
holes as St. Augustine's objects can survive such a dramatic event.
Therefore, we conclude that it is possible to have some amount of black hole
population, moving relativistically with respect to the nearby Hubble flow
at high redshifts, as a part of the Big Bang initial conditions. If the amount
of such St. Augustine's objects is small, they cannot significantly change the
dynamics of the universe and spoil the successful predictions of the standard
cosmological model.

Let us sketch, from the Milne model perspective, how St. Augustine's objects
in the role of quasars can offer an explanation of some above mentioned
mysteries.

Let us begin with the time non-dilation mystery. It is expected that the most
radiation from quasars comes from the ambient matter heated by shock waves
when St. Augustine's black hole moves relativistically through this matter.
Therefore, the associated red-shifts will be truly cosmological as originated
from the matter which participates in the Hubble flow. On the other hand, any
internal variability related to the black hole itself (for example, due to
instabilities in the accretion disc) will not be time-dilated if the
St. Augustine's object is nearly motionless with respect to us in our
\mbox{private space.}

The mystery related to the  origin of supermassive black holes also becomes
less acute. Indeed, St. Augustine's black holes might already have been
sufficiently massive upon entering our universe. Moreover, initially
St. Augustine's black hole moves ultrarelativistically through ambient matter
and extremely high value of its gamma-factor drives it into the fast
growing mode \cite{99}.

Such high gamma-factors are considered as unrealistic for primordial black
holes in the early \mbox{universe \cite{99}}, but for St. Augustine's black 
holes this is just what is expected most naturally from anthropic principle. 
Indeed, if we assume that the relativistic velocity of the St. Augustine's
black hole with respect to ambient Hubble flow matter is essential for
triggering the quasar activity, an immediate anthropic principle type
consequence would be that we are not a typical observer but rather a special
one, namely the one which is nearly at rest with respect to potential
St. Augustine's objects population, because otherwise the existence of nearby
quasars would make development of intelligent life impossible. At that,
of course, we have implicitly assumed that St. Augustine's objects population
indeed define a reasonable standard of rest. To make sure that this is
a plausible supposition, imagine an analog FRW space-time created in
a hydrodynamical laboratory experiment, as advocated in \cite{98}. Then the
vessel in which an expanding relativistic fluid mimics the FRW universe
provides a natural standard of rest and various residual gas molecules, as
candidate St. Augustine's objects for this simulated FRW universe,
approximately can be considered as nearly at rest with respect to the vessel
from the point of view of participants of the relativistic flow.

As relative velocities of St. Augustine's black holes with respect to ambient
Hubble flow matter increase with redshift, it is expected that high redshift
quasars will be more luminous than low redshift quasars. Therefore, the
observed evolution of quasar luminosity over redshifts is no longer
mysterious but finds its explanation if quasars are related to St. Augustine's
objects.

The shock wave, which surely accompanies the motion of the  St. Augustine's
black holes through ambient matter, is expected to trigger intense star
formation in this ambient matter and hence increase the expected supernovae
rates. Thus, it becomes possible to explain the appearance of unusually young
chemically mature galaxies around high-redshift quasars.

\begin{figure}[ht]
 \begin{center}
    \includegraphics[scale=0.7]{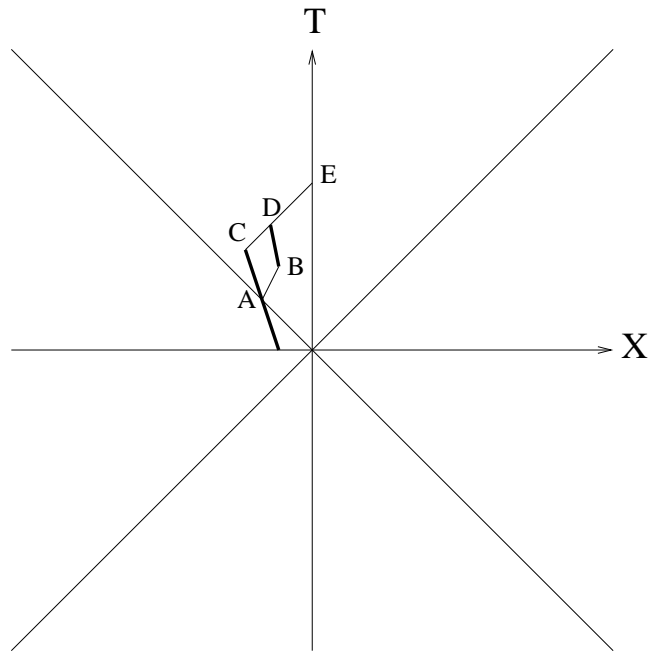}
  \end{center}
\caption{A schematic mechanism explaining correlations of objects that have
vastly different redshifts. See text for details.}
\label{fig5}
\end{figure}

Probably, shock waves caused by St. Augustine's objects can give a clue also
to the most mysterious of paradoxes related to quasars: alleged correlations
of objects that have vastly different redshifts. Schematically this is
explained on Figure ~\ref{fig5}.
At a space-time point $A$ a St. Augustine's object enters into the Milne
universe and at the present time we, situated at the point $E$, observe
the light emitted by the corresponding quasar at point $C$. The accompanying
shock wave propagates along the world line $AB$ and at a point $B$ triggers
the formation of the seed of future galaxy which sends to us the light from
the point $D$ when it is already formed. As a result, we observe a distant
quasar and much closer galaxy aligned, and this alignment is really not by
a chance, but due to above described fancy causal relation between them.

It can be expected that the shock wave will give birth not to one but to
several galaxies along its world line. Therefore, we expect chains of galaxies
with different redshifts all aligned with a quasar. Remarkably, according to
Arp \cite{42}, there are actually many examples of such chains of galaxies
throughout extragalactic space. In fact this effect is so pronounced that
it led Victor Ambartsumian in 1958 to the exotic idea that new galaxies are
formed through the ejection from older active \mbox{galaxies \cite{44}}. 
He came to this idea by simply looking at pictures of galaxies. Eight years 
later Arp came to the same conclusion by inspecting much more detailed and
significantly better quality images of peculiar galaxies \cite{100}.
Ambartsumian presented his conclusions at the prestigious Solvay conference.
However, ``this select group of the best known scientists in the
world had either been completely baffled or laughed privately at these crazy
ideas'' \cite{100}. Interestingly enough, the paradigm of St. Augustine's
objects reconciles Ambartsumian's and Arp's exotic idea that galaxies beget
galaxies with the standard cosmological interpretation of redshifts in
a manner which does not undermine the standard theory of galaxy formation for
the bulk of galaxy population, if the number of the St. Augustine's objects
is \mbox{small enough.}

As the last bonus, the paradigm of St. Augustine's objects also explains why
the observation of an apparent superluminal motion in quasars is not
a relatively rare phenomenon. In public space, St. Augustine's objects move
relativistically nearly towards us so that the associated blue-shift nearly
compensates the cosmological redshift for any internal variation of intensity.
As a by product, relative motion of any quasar-related two radio emitting
blobs will be with a  high probability radial and hence can lead to the
apparent superluminal motion.
	
\section{Concluding Remarks}

We have heavily used the Milne model in this article to justify our
suppositions. It is time to ask what are the relations of this model with
reality. We introduced the Milne metric as a metric of an empty FRW universe.
The present density of matter is rather low in terms of the critical density.
Therefore, probably, it is not surprising that the Milne metric describes
some aspects of the present day universe quite satisfactorily. Namely,
curiously the Milne model explains supernovae observations rather well without
requiring any dark energy \cite{101,101A}. The standard concordance
$\Lambda$CDM model with positive cosmological constant, of course, provides
a better fit and, as expected, an empty universe is not its viable
alternative. Nevertheless, ``the Milne model has great pedagogical value,
elucidating the kinematic aspect of the universe's expansion'' \cite{101A}.

However, in fact, Milne's aspiration was significantly more ambitious than
just providing late time approximate description of FRW cosmology in terms of
special relativity. Milne believed that some very general rational and
at the same time aesthetic principles were at the foundations of the observed
order and regularities in the universe. His approach, named kinematic
relativity \cite{63}, was conceived as a rival theory to the standard
relativistic cosmology. Milne's aim was ``to deduce as much as possible
merely from the cosmological principle and the basic properties of space,
time and the propagation of \mbox{light'' \cite{102}} revealed by special 
relativity.

Milne's unorthodox methodology triggered a heated debate involving the most
eminent cosmologists of the time \cite{103,104,105}. When the kinematic
relativity was first developed ``it met great hostility and was
criticized very severely, often unjustly, and sometimes frivolously''
\cite{102}.

In retrospect, the influence of Milne on the development of modern cosmology
was very substantial and he inspired others, in particular Robertson and
Walker, in shaping fundamental concepts of modern cosmology
\cite{103,104,105}. However, as for the kinematic relativity, very little
work has been done on it since the death of Milne and the theory has been
left ``in a curiously unfinished state'' \cite{102}. Walker showed
\cite{106,107} that Milne's cosmological construction, if described in terms
of geometry, was fundamentally different from that of general relativity and,
in general, such a description requires Finsler geometry, a generalization
of Riemannian geometry (for applications of Finsler geometry see, for example,
\cite{108,109}). Although we feel that the Finslerian perspective of the
Milne model was not sufficiently explored, at present the Milne model cannot
be considered as a viable alternative to the standard relativistic cosmology,
of course. Therefore, in the spirit of this paper we can use it as only
a source of inspiration and not a solid ground to justify
the introduction of the St. Augustine's objects. Can we find a support for
them in more mainstream theories?

In non-empty FRW cosmological models we cannot define global inertial frames
and thus, we cannot introduce private time like Milne did. However,
a plausible generalization of the concept of private time can be envisaged
in the general case as well thanks to the fact that Weyl Tensor for all FRW
Cosmological metrics vanishes and thus all FRW space-times are conformally
flat \cite{110,111}. Suitably defined conformal time can serve as a substitute
of  private time. For example, in the case of the flat FRW metric
we can introduce conformal time-coordinate $T$ and conformal radial
coordinate $R$ in the following way \cite{111}
\begin{equation}
T=\frac{\eta}{\eta^2-\psi^2},\;\;\; R=\frac{\psi}{\eta^2-\psi^2}
\label{eq111}
\end{equation}
where
\begin{equation}
\eta=\int\limits_{t_0}^t\frac{dt}{a(t)}
\label{eq112}
\end{equation}
with arbitrary $t_0$. Substituting the inverse transformations
\begin{equation}
\eta=\frac{T}{T^2-R^2},\;\;\; \psi=\frac{R}{T^2-R^2}
\label{eq113}
\end{equation}
into Eq. (\ref{eq4}) and taking into account that according to 
Eq. (\ref{eq112}) $dt=a(t)d\eta$, we find that the metric Eq. 
(\ref{eq4}) (with $k=0$) indeed takes the conformally flat form:
\begin{equation}
ds^2=\frac{a(t(T,R))^2}{(T^2-R^2)^2}\left[dT^2-dR^2-R^2(d\theta^2+
\sin^2{\theta}d\phi^2)\right]
\label{eq114}
\end{equation}

According to Eq. (\ref{eq111}), $0<\psi<\eta$ region is mapped onto the 
region $0<R<T$, while the  region $0<\eta<\psi$---onto the region $R<T<0$. 
As in the case of private space and time in the Milne model, the FRW universe
coordinates occupy only a quarter of the $(T,R)$ plane and this circumstance
hints at a possibility that the FRW universe might be just a part of some
bigger entity. Let us note, however, that the conformal coordinates
considered are not exact analogues of the Milne's private time and
private space. In particular, transformations Eq. (\ref{eq111}) are 
singular at the light-cone $\eta=\psi$.

Another analogue of the Milne foliation is the Rindler's foliation of
a non-empty FRW \mbox{space-time \cite{65}} by spacelike slices of finite 
volume. Rindler showed that, just like the Milne universe, ``every open 
big-bang FRW universe can be regarded as an expanding finite-volume ball of 
matter, springing from zero volume'' \cite{65}. This feature of FRW space-time
becomes particularly clear from the multi-dimensional perspective \cite{65}.
It is well known that any four-dimensional space-time can be locally
embedded (immersed) in a ten-dimensional flat pseudo-Euclidean space
\cite{112}. Because of the high degree of symmetry of the FRW space-time,
its embedding is possible even in a five-dimensional pseudo-Euclidean space
\cite{113}. Namely, the embedding of a spatially flat FRW spacetime
\begin{equation}
ds^2=dt^2-a^2(t)(dx^2+dy^2+dz^2)
\label{eq115}
\end{equation}
into the five-dimensional Minkowski space is defined by relations \cite{113}:
\begin{eqnarray} &&
T=\frac{a(t)}{2\alpha}(\alpha^2+r^2)+\frac{1}{2\alpha}\int\limits_{t_0}^t
\frac{d\tau}{\dot{a}(\tau)},\;\; X=a(t)x,\;\; Y=a(t)y \nonumber \\ &&
Z=a(t)z,\;\; Z_5=\frac{a(t)}{2\alpha}(\alpha^2-r^2)-
\frac{1}{2\alpha}\int\limits_{t_0}^t\frac{d\tau}{\dot{a}(\tau)}=\alpha a(t)-
T
\label{eq116}
\end{eqnarray}
where $r^2=x^2+y^2+x^2$ and $\alpha$ is a constant with the
dimension of length.

It follows from Eq. (\ref{eq116}) that \cite{113}
\begin{equation}
T^2-X^2-Y^2-Z^2-Z_5^2=a(t)\int\limits_{t_0}^t\frac{d\tau}{\dot{a}(\tau)}
\label{eq117}
\end{equation}

In the right part of Eq. (\ref{eq117}), the cosmic time $t$ can be
considered as a function of $T+Z_5$ because, according to Eq. 
(\ref{eq116}),
\begin{equation}
T+Z_5=\alpha a(t)
\label{eq118}
\end{equation}

Therefore, Eq.s (\ref{eq117}) and (\ref{eq118}) define a hyper-surface
in the five-dimensional Minkowski space. For example, in the radiation
dominated universe $a(t)=\sqrt{t/\tau}$ with some $\tau$, and the
Eq. (\ref{eq117}), defining the hyper-surface, takes the form (in the case
$t_0=0$)
\begin{equation}
T^2-X^2-Y^2-Z^2-Z_5^2=\frac{4\tau^2}{3}\left(\frac{T+Z_5}{\alpha}\right)^4
\label{eq119}
\end{equation}

In fact, the FRW space-time corresponds to only a portion of this
hyper-surface, because according to Eq. (\ref{eq118}) we must have 
$T+Z_5\ge 0$.

As we see, a multi-dimensional perspective allows to consider a non-empty
FRW universe as a part of a bigger entity and the overall picture is similar
to the one in the Milne model that led us to the St. Augustine's objects
hypothesis.

It is worth mentioning that extra-dimensional models were very popular in the 
last decade.
In such models, an ambient space is not the Minkowski space but has more
complex geometry. The universe as a thin shell expanding in a five-dimensional
space with so called non-factorizable warped geometry was first considered
by Gogberashvili in \cite{114}, and such models with warped extra dimensions
became very popular after the Randall and Sundrum paper \cite{115}.
Other popular extra-dimensional models are Kaluza-Klein type models
with compactified extra-dimensions \cite{116}. It is assumed that all Standard
Model particles are confined to the four-dimensional hyper-surface, a brane,
in a higher dimensional space. Gravity, on the contrary, can freely propagate
in the bulk thus explaining the weakness of the gravitational interactions
between the brane confined particles by accompanying spreading of
gravitational flux into the large volume of the extra dimensions. In our
terminology, gravitons are St. Augustine's objects in these models, and who
knows what other type of St. Augustine's objects are lurking in the bulk.

Gravitons also play the role of St. Augustine's objects in Penrose's
conformal cyclic \mbox{cosmology \cite{117}}. Gravitational waves emitted in
close encounters between super-massive black holes in the last contracting
phase of the previous aeon can propagate into the next aeon giving
rise to fancy concentric circular rings with slightly different temperature
on the Cosmic microwave background sky \cite{117}.

Finally, let us mention the so called Ekpyrotic cosmology in which the Big Bang
is associated not with a singularity but with a collision of two branes
in the extra-dimensional bulk \cite{118}. As a result,  cyclic model of the
universe can be constructed in which there is an endless sequence of cosmic
epochs of expansion and contraction \cite{119}. Black holes as possible
St. Augustine's objects in such  cyclic scenarios, surviving a bounce of
the Big Crunch-Big Bang transition, are discussed in \cite{120}.
Interestingly, modern attempts to discuss sailing through the big crunch-big
bang transition involves the (compactified) Milne universe in an essential
way \cite{91,121}. Milne's influence on the modern cosmology is ongoing!

\vspace{32pt}
\noindent{\textbf{Acknowledgements}}\vspace{12pt}

The work of Z.K.S. is supported by the Ministry of Education and Science of
the Russian Federation and in part by Russian Federation President
Grant for the support of scientific schools  NSh-2479.2014.2 and by
RFBR grant 13-02-00418-a.


\begin{thebibliography}{------}

\bibitem{E}
Tyutchev, F.I. A Spring Storm. 1828. Available online: \\ 
http://www.ruthenia.ru/tiutcheviana/publications/trans/springstorm.html 
(accessed on 28 September 2015).

\bibitem{15}
Hawkins, M.R.S.
On time dilation in quasar light curves.
\emph{Mon.\ Not.\ Roy.\ Astron.\ Soc.\ } \textbf{2010}, {\textit {405}}, 
1940--1946

\bibitem{I-1}
Veltman, M.
\emph{Diagrammatica}; Cambridge University Press: Cambridge, UK, 1995.

\bibitem{1}
Lineweaver, C.H.; Davis, T.M.
Misconceptions about the Big Bang.
\emph{Sci.\ Am.\ } \textbf{2005}, {\textit{ 292}}, 36--45.

\bibitem{1-A}
Francis, M.J.; Barnes, L.A.; James, J.B.; Lewis, G.F.
Expanding Space: The Root of all Evil?
\emph{Publ.\ Astron.\ Soc.\ Austral.\ } \textbf{2007}, {\textit {24}}, 95--102.

\bibitem{1-B}
Peacock, J.A.
A diatribe on expanding space. \textbf{2008},
{arXiv:0809.4573 [astro-ph].}

\bibitem{1-C}
Braeck, S.; Elgar\o y, \O.
A physical interpretation of Hubble's law and the cosmological redshift
from the perspective of a static observer.
\emph{Gen.\ Rel.\ Grav.\ } \textbf{2012}, {\textit {44}}, 2603--2610.

\bibitem{1-D}
Cook, R.J.; Burns, M.S.
Interpretation of the Cosmological Metric.
\emph{Am.\ J.\ Phys.\ } \textbf{2009}, {\textit {77}}, 59--66

\bibitem{1-E}
Chodorowski, M.
A direct consequence of the expansion of space?
\emph{Mon.\ Not.\ Roy.\ Astron.\ Soc.\ } \textbf{2007}, {\textit {378}}, 
239--244

\bibitem{1-F}
Melia, F.
Cosmological redshift in Friedmann-Robertson-Walker metrics with
constant space-time curvature.
\emph{Mon.\ Not.\ Roy.\ Astron.\ Soc.\ }\textbf{ 2012}, {\textit {422}}, 
1418--1424.

\bibitem{1-G}
Gr\o n, \O.; Elgar\o y, \O.
Is space expanding in the Friedmann universe models?
\emph{Am.\ J.\ Phys.\ } \textbf{2007}, {\textit {75}}, 151--157

\bibitem{1-H}
Roukema, B.F.
There was movement that was stationary, for the four-velocity
had passed around.
\emph{Mon.\ Not.\ Roy.\ Astron.\ Soc.\ } \textbf{2010}, {\textit {404}}, 
318--324.

\bibitem{1-I}
Hartnett, J.G.
Is the Universe really expanding? \textbf{2011},
{arXiv:1107.2485 [physics.gen-ph].}

\bibitem{1-J}
Lopez-Corredoira, M.
Tests for the Expansion of the Universe. \textbf{2015},
{arXiv:1501.01487 [astro-ph.CO].}

\bibitem{1-K}
Mitra, A.
When can an ``Expanding Universe'' look ``Static'' and vice versa: A
comprehensive study.
\emph{Int.\ J.\ Mod.\ Phys.\ D} \textbf{2015}, {\textit {24}}, 1550032.

\bibitem{2}
Minkowski, H.
Space and Time.
In \emph{The Principle of Relativity: A Collection of Original Memoirs on
the Special and General Theory of Relativity}; Lorentz, H.A.; Einstein, A.; 
Minkow\-ski, H.; \mbox{Weyl, H.;} Eds.; Dover Publications: New York, USA, 
1952; p. 75.

\bibitem{3}
Wigner, E.P.
Relativistic invariance and quantum phenomena.
\emph{Rev.\ Mod.\ Phys.\ } \textbf{1957}, {\textit {29}}, 255--268.

\bibitem{4}
Carroll, S.M.
Lecture notes on general relativity. \textbf{1997},
{arXiv:gr-qc/9712019.}

\bibitem{5}
Giulini, D.
Does cosmological expansion affect local physics? \textbf{2013},
{arXiv:1306.0374 [gr-qc].}

\bibitem{6}
Rindler, W.
\emph{Relativity: Special, General and Cosmological};
Oxford University Press: Oxford, \mbox{UK, 2001.}

\bibitem{7}
Herranz, F.J.; Ortega, R.; Santander, M.
Trigonometry of spacetimes: A new self-dual approach to a
curvature/signature (in)dependent trigonometry.
\emph{J.\ Phys.\ A} \textbf{2000}, {\textit {33}}, 4525--4551.

\bibitem{8}
Ballesteros, A.; Herranz, F.J.; del Olmo, M.A.; Santander, M.
Quantum structure of the motion groups of the two-dimensional
Cayley-Klein geometries.
\emph{J.\ Phys.\ A} \textbf{1993}, {\textit {26}}, 5801--5823.

\bibitem{9}
Narlikar, J.V.
\emph{An Introduction to Cosmology}; Cambridge University Press:
Cambridge, UK, 1993.

\bibitem{10}
``The great tragedy of Science: the slaying of a beautiful hypothesis by an
ugly fact'' (Thomas Henry Huxley, 1870). Qouted in \emph{The Yale Book of 
Quotations}; Shapiro, F.~R.; Epstein, J.; Eds.; Yale University Press: 
London, UK, 2006; p. 379.

\bibitem{11}
Zee, A.
\emph{Einstein Gravity in a Nutshell}; Princeton University Press:
Princeton, NJ, USA, 2013.

\bibitem{12}
Moschella, U.
The de Sitter and anti-de Sitter Sightseeing Tour. In \emph{Einstein, 
1905-2005, Poincar\'e Seminar 2005}; Damour, T.; Darrigol, O.; Duplantier,
B.; Rivasseau, V.; Eds.; \emph{Progress in Mathematical Physics,
Volume 47}; Birkh\"{a}user Verlag: Basel, Switzerland, 2006; pp. 120--133.

\bibitem{13}
Blondin, S.; Davis, T.~M.; Krisciunas, K.; Schmidt, B.~P.; Sollerman, J.; 
Wood-Vasey W.~M.; Becker, A.~C.; Challis, P.; Clocchiatti, A.; Damke, G.;
{\it et al.} 
Time Dilation in Type Ia Supernova Spectra at High Redshift.
\emph{Astrophys.\ J.\ } \textbf{2008}, {\textit {682}}, 724--736

\bibitem{14}
Wilson, O.C.
Possible Applications of Supernovae to the Study of the Nebular Red
Shifts.
\emph{Astrophys.\ J.\ }\textbf{ 1939}, {\textit {90}}, 634--636.

\bibitem{14A}
Crawford, D.F.
No Evidence of Time Dilation in Gamma-Ray Burst Data. \textbf{2009},
{arXiv:0901.4169 [astro-ph.CO].}

\bibitem{14B}
Kocevski, D.; Petrosian, V.
On The Lack of Time Dilation Signatures in Gamma-ray Burst Light Curves.
\emph{Astrophys.\ J.} \textbf{2013}, {\textit {765}}, 116.

\bibitem{14C}
Littlejohns, O.M.; Butler, N.R.
Investigating signatures of cosmological time dilation in duration
measures of prompt gamma-ray burst light curves.
\emph{Mon.\ Not.\ Roy.\ Astron.\ Soc.} \textbf{2014}, {\textit {444}}, 
3948--3960.

\bibitem{14D}
Zhang, F.W.; Fan, Y.Z.; Shao, L.; Wei, D.M.
Cosmological Time Dilation in Durations of Swift Long Gamma-Ray Bursts.
\emph{Astrophys.\ J.} \textbf{2013}, {\textit {778}}, L11

\bibitem{15A}
Ellis, G.F.R.
Contributions of K. G\"{o}del to relativity and cosmology.
\emph{Lect. Notes  Log.} \textbf{1996}, {\textit 6}, 34--49.

\bibitem{15AA}
Lanczos, K.
On a Stationary Cosmology in the Sense of Einstein's Theory of
Gravitation.
\emph{Gen.\ Rel.\ Grav.}\textbf{ 1997}, {\textit {29}}, 363--399.

\bibitem{15AAA}
Van Stockum, W.J.
The gravitational field of a distribution of particles rotating around
an axis of symmetry.
\emph{Proc.\ Roy.\ Soc.\ Edinburgh} \textbf{1937}, {\textit {57}}, 135--154.

\bibitem{15B}
Hawking, S.
The Existence of cosmic time functions.
\emph{Proc.\ Roy.\ Soc.\ Lond.\ A} \textbf{1968}, {\textit {308}}, 433--435.

\bibitem{15BB}
Geroch, R.P.
General relativity in the large.
\emph{Gen.\ Rel.\ Grav.} \textbf{1971}, {\textit 2}, 61--74.

\bibitem{15C}
Geroch, R.; Horowitz, G.T.
Global structure of spacetimes. In  \emph{General Relativity: An Einstein 
Centenary Survey}; Hawking, S.; Isreal, W.; Eds.; Cambridge University Press: 
Cambridge, UK, 1979; pp. 212--293.

\bibitem{15D}
Minguzzi, E.
On the global existence of time.
\emph{Int.\ J.\ Mod.\ Phys.\ D} \textbf{2009}, {\textit {18}}, 2135--2144.

\bibitem{15E}
Weyl, H.
\emph{Raum, Zeit, Materie}, 5th ed.; Springer: Berlin, Germany, 1923.

\bibitem{15F}
Rugh, S.E.; Zinkernagel, H.
Weyl's principle, cosmic time and quantum fundamentalism. \textbf{2010},
{arXiv:1006.5848 [gr-qc].}

\bibitem{16}
Kellermann, K.I.
The Discovery of Quasars.
\emph{Bull.\ Astron.\ Soc.\ India} \textbf{2013}, {\textit {41}}, 1--17.

\bibitem{17}
D'Onofrio, M.; Marziani, P.; Sulentic, J.~W.; Collin, S.; Setti, G.; 
Gaskell, M.; Wampler, J.; Elvis, M.; Pronik, I.; Pronik, V.; {\it et al.} 
Quasars in the Life of Astronomers. In
\emph{Fifty Years of Quasars: From Early Observations and Ideas to Future
Research}; D'Onofrio, M.; Marziani, P.; Sulentic, J.~W.; Eds.; Springer: 
Heidelberg, Germany, 2012; pp. 11--90.

\bibitem{18}
Zel'dovich, Ya.B.
The Fate of a Star and the Evolution of Gravitational Energy Upon
Accretion.
\emph{Sov.\ Phys.\ Dokl.\ } \textbf{1964}, {\textit 9}, 195--197.

\bibitem{19}
Salpeter, E.E.
Accretion of interstellar matter by massive objects.
\emph{Astrophys.\ J.\ } \textbf{1964}, {\textit {140}}, 796--800.

\bibitem{20}
Lynden-Bell, D.
Galactic nuclei as collapsed old quasars.
\emph{Nature} \textbf{1969}, {\textit {223}}, 690--694.

\bibitem{21}
Ojha, R.
Parsec-scale structure of quasars: dawn of the golden age? \textbf{2013}, 
{arXiv:1310.0875 [astro-ph.HE].}

\bibitem{22}
Antonucci, R.
Astrophysics: Quasars still defy explanation.
\emph{Nature} \textbf{2013}, {\textit {495}}, 165--167.

\bibitem{23}
Lopez-Corredoira, M.
Pending problems in QSOs. \textbf{2009},
{arXiv:0910.4297 [astro-ph.CO].}

\bibitem{24}
Kocsis, B.; Loeb, A.
Menus for Feeding Black Holes. \textbf{2013},
{arXiv:1310.0815 [astro-ph.CO].}

\bibitem{25}
Frank, J.; King, A.; Raine, D.
\emph{Accretion Power in Astrophysics}; Cambridge University Press:
Cambridge, UK, 2002.

\bibitem{26}
Peterson, B.M.
\emph{An Introduction to Active Galactic Nuclei}; Cambridge University Press:
Cam\-bridge, UK,  1997.

\bibitem{27}
Shen, Y.
The Mass of Quasars.
\emph{Bull.\ Astron.\ Soc.\ India} \textbf{2013}, {\textit {41}}, 61--115

\bibitem{28}
Paczy\'{n}sky, B.; Wiita, P.J.
Thick accretion disks and supercritical luminosities.
\emph{Astron.\ Astrophys.\ } \textbf{1980}, {\textit {88}}, 23--31.

\bibitem{29}
Abramowicz, M.A.
The Paczynski-Wiita potential: A step-by-step ``derivation''.
\emph{Astron.\ Astrophys.\ } \textbf{2009}, {\textit {500}}, 213--214.

\bibitem{30}
Hobson, M.P.; Efstathiou, G.P.; Lasenby, A.N.
\emph{General Relativity: An Introduction for Physicists}; Cambridge
University Press: Cambridge, UK, 2006.

\bibitem{31}
Alexander, D.M.; Hickox, R.C.
What Drives the Growth of Black Holes?
\emph{New Astron.\ Rev.\ } \textbf{2012}, {\textit {56}}, 93--121.

\bibitem{32}
Narlikar, J.~V. Alternative Views and Ideas. In \cite{17}; p. 77.

\bibitem{33} 
Finkelstein, S.~L.; Papovich, C.;  Dickinson, M.; Song, M.;  Tilvi, V.;
Koekemoer, A.~M.; Finkelstein, K.~D.; Mobasher, B.; Ferguson, H.~C.;
Giavalisco. M.; {\it et al.}
A Rapidly Star-forming Galaxy 700 Million Years After the Big Bang
at z=7.51. \textbf{2013},
{arXiv:1310.6031 [astro-ph.CO].}

\bibitem{34}
Jiang, L.; Fan, J.; Brandt, W.~N.; Carilli, C.~L.; Egami, E.; Hines, D.~C.;
Kurk, J.~D.; Richards, G.~T.; Shen, Y.; Strauss, M.~A.; {\it et al.}
Dust-free quasars in the early Universe.
\emph{Nature} \textbf{2010}, {\textit {46}}, 380--383.

\bibitem{35}
Jain, D.; Dev, A.
Age of high redshift objects - a litmus test for the dark energy
models.
\emph{Phys.\ Lett.\ B} \textbf{2006}, {\textit {633}}, 436--440.

\bibitem{36}
Friaca, A.; Alcaniz, J.; Lima, J.A.S.
An Old quasar in a young dark energy-dominated Universe?
\emph{Mon.\ Not.\ Roy.\ Astron.\ Soc.\ } \textbf{2005}, {\textit {362}}, 
1295--1300.

\bibitem{37}
Kobayashi, C.; Umeda, H.; Nomoto, K.; Tominaga, N.; Ohkubo, T.
Galactic chemical evolution: Carbon through Zinc.
\emph{Astrophys.\ J.\ } \textbf{2006}, {\textit {653}}, 1145--1171.

\bibitem{38}
Kobayashi, C.; Nomoto, K.
The Role of Type Ia Supernovae in Chemical Evolution I: Lifetime
of Type Ia Supernovae and Galactic Supernova Rates.
\emph{Astrophys.\ J.\ } \textbf{2009}, {\textit {707}}, 1466--1484.

\bibitem{39}
Wang, S.; Li, X.-D.; Li, M.
Revisit of cosmic age problem.
\emph{Phys.\ Rev.\ D} \textbf{2010}, {\textit {82}}, 103006.

\bibitem{40}
Feige, B.
Elliptic integrals for cosmological constant cosmologies.
\emph{Astron.\ Nachr.\ } \textbf{1992}, {\textit {313}}, 139--163.

\bibitem{41}
Thomas, R.C.; Kantowski, R.
Age redshift relation for standard cosmology.
\emph{Phys.\ Rev.\ D} \textbf{2000}, \mbox{{\textit {62}}, 103507.}

\bibitem{42}
Arp, H.
\emph{Quasars, Redshifts and Controversies}; Cambridge University Press:
Cambridge, UK,~1988.

\bibitem{43}
Kembhavi, A.K.; Narlikar, J.V.
\emph{Quasars and Active Galactic Nuclei: An Introduction}; Cambridge
University Press: Cambridge, UK, 1999.

\bibitem{44}
Lopez-Corredoira, M.
Apparent discordant redshift QSO-galaxy associations. \textbf{2009},
{arXiv:0901.4534 [astro-ph.CO].}

\bibitem{45}
Lopez-Corredoira, M.; Gutierrez, C.M.
Two emission line objects with z > 0.2 in the optical filament
apparently connecting the Seyfert galaxy NGC 7603 to its companion.
\emph{Astron.\ Astrophys.\ } \textbf{2002}, {\textit {390}}, L15--L18.

\bibitem{46}
Lopez-Corredoira, M.; Gutierrez, C.M.
The Field surrounding NGC 7603: Cosmological or non-cosmological
redshifts?
\emph{Astron.\ Astrophys.\ } \textbf{2004}, {\textit {421}}, 407--423.

\bibitem{47}
Rees, M.J.
Appearance of Relativistically Expanding Radio Sources.
\emph{Nature} \textbf{1966}, {\textit {211}}, 468--470.

\bibitem{48A}
Blandford, R.D.; McKee, C.F.; Rees, M.J.
Super-luminal expansion in extragalactic radio sources.
\emph{Nature} \textbf{1977}, {\textit {267}}, 211--216.

\bibitem{48}
Falla, D.F.; Floyd, M.J.
Superluminal motion in astronomy.
\emph{Eur.\ J.\ Phys.\ } \textbf{2002}, {\textit {23}}, 69--81.

\bibitem{49}
Kellermann, K.I.; Lister, M.~L.; Homan, D.~C.; Vermeulen, R.~C.; Cohen, M.~H.;
Ros, E.;  Kadler, M.; Zensus, J.~A.; Kovalev, Y.~Y.
Sub-milliarcsecond imaging of quasars and Active Galactic Nuclei
3. Kinematics of parsec---scale radio jets.
\emph{Astrophys.\ J.\ } \textbf{2004}, {\textit {609}}, 539--563.

\bibitem{50}
Belchenko, Yu.~I.; Gilev, E.~A.; Silagadze, Z.K.
\emph{Problems in Mechanics of Particles and Bodies}; RCD: Moscow-Izhevsk,
Russia, 2008;  problem 2.6. (In Russian)

\bibitem{51}
Chodorowski, M.J.
Superluminal apparent motions in distant radio sources.
\emph{Am.\ J.\ Phys.\ } \textbf{2005}, {\textit {73}}, 639--643.

\bibitem{52}
Narlikar, J.V.; Chitre, S.M.
Faster-than-Light Motion in Quasars.
\emph{J.\ Astrophys.\ Astron.\ } \textbf{1984}, {\textit {5}}, 495--506.

\bibitem{53}
Lopez-Corredoira, M.; Perucho, M.
Kinetic power of quasars and statistical excess of MOJAVE superluminal
motions.
\emph{Astron.\ Astrophys.\ } \textbf{2012}, {\textit {544}}, A56.

\bibitem{54}
Ryle, M.; Longair, M.S.
A Possible Method for Investigating the Evolution of Radio Galaxies.
\emph{Mon.\ Not.\ Roy.\ Astron.\ Soc.\ } \textbf{1967}, {\textit {136}}, 
123--140.

\bibitem{55}
Urry, C.M.; Padovani, P.
Unified schemes for radio-loud active galactic nuclei.
\emph{Publ.\ Astron.\ Soc.\ Pac.\ } \textbf{1995}, {\textit {107}}, 803--845.

\bibitem{56}
Rybicki, G.B.; Lightman, A.P.
\emph{Radiative Processes in Astrophysics}; Wiley-VCH: Weinheim, Germany, 2004.

\bibitem{56L}
Landau, L.D.; Lifshitz, E.M.
\emph{Course of Theoretical Physics, Vol. 2, The Classical Theory of Fields};
Pergamon Press: Oxford, UK, 1975.

\bibitem{56A}
Debbasch, F.; Rivet, J.-P.; van Leeuwen, W.
Invariance of the relativistic one-particle distribution function.
\emph{Phys. A} \textbf{2001}, {\textit {301}}, 181--195.

\bibitem{56B}
Treumann, R.A.; Nakamura, R.; Baumjohann, W.
Relativistic transformation of phase-space distributions.
\emph{Ann.\ Geophys.\ } \textbf{2011}, {\textit {29}}, 1259--1265.

\bibitem{57}
Castor, J.I.
\emph{Radiation Hydrodynamics}; Cambridge University Press: Cambridge, UK, 
2004.

\bibitem{58}
Bradt, H.
\emph{Astrophysics Processes: The Physics of Astronomical Phenomena};
Cambridge University Press: Cambridge, UK, 2008.

\bibitem{59}
Lind, K.R.; Blandford, R.D.
Semidynamical models of radio jets---Relativistic beaming and source
counts.
\emph{Astrophys.\ J.\ } \textbf{1985}, {\textit {295}}, 358--367.

\bibitem{60}
Bell, M.B.
Doppler Boosting May Have Played No Significant Role in the Finding
Surveys of Radio-Loud Quasars.
\emph{Int.\ J.\ Astron.\ Astrophys.\ } \textbf{2012}, {\textit 2}, 52--61.

\bibitem{JB}
Parker, R.H.; Wolnizer, P.W.; Nobes, C.; (Eds.)
\emph{Readings in True and Fair}; Routledge: New York, NY, USA, 1996; p. 74. 
(Another version of the quotation can be found in R.A.Muller,
\emph{Nemesis for Nemesis?}
Available online: http://muller.lbl.gov/pages/nemfornem.htm, accessed on 
21.09.2015).

\bibitem{61} Milne, E.A.
World-Structure and the Expansion of the Universe.
\emph{Z.\ Astrophys.\ } \textbf{1933}, {\textit 6}, 1--95.

\bibitem{62}
Milne, E.A.
\emph{Relativity, Gravitation and World-Structure}; Oxford
University Press: London, \mbox{UK, 1935.}

\bibitem{63}
Milne, E.A.
\emph{Kinematic Relativity}; Oxford University Press: London, UK, 1948.

\bibitem{64}
Milne, E.A.
\emph{Modern Cosmology and the Christian Idea of God}; Clarendon Press:
Oxford, \mbox{UK, 1952.}

\bibitem{65}
Rindler, W.
Finite foliations of open FRW universes and the point-like big bang.
\emph{Phys.\ Lett.\ A} \textbf{2000}, {\textit {276}}, 52--58.

\bibitem{66}
Sexl, R.U.; Urbantke, H.K.
\emph{Relativity, Groups, Particles: Special Relativity and Relativistic
Symmetry in Field and Particle Physics}; Springer: Vienna, Austria, 2001; 
p. 38.

\bibitem{67}
Hearn, A.C.
\emph{Reduce User's Manual}; Rand Corporation: Santa Monica, CA, USA, 1989.

\bibitem{68}
Silberstein, L.
\emph{The Theory of Relativity}; MacMillan: London, UK, 1914; p. 169.

\bibitem{69}
Ungar, A.A.
\emph{Beyond the Einstein Addition Law and Its Gyroscopic
Thomas Precession: The Theory of Gyrogroups and Gyrovector Spaces};
Kluwer Academic Publishers: New York, NY, USA, 2002; p. 18.

\bibitem{70}
Robertson, H.P.
On E.A.Milne's Theory of World Structure.
\emph{Z.\ Astrophys.\ } \textbf{1933}, {\textit 7}, 153--166.

\bibitem{71}
Jammer, M.
\emph{Concepts of Simultaneity: From Antiquity to Einstein and Beyond};
Johns Hopkins University Press: Baltimore, MA, USA, 2006.

\bibitem{72}
Gr{\o}n, \O.
Big bang in a universe with infinite extension.
\emph{Eur.\ J.\ Phys.\ } \textbf{2006}, {\textit {27}}, 561--565.

\bibitem{73}
Dirac, P.A.M.
Forms of Relativistic Dynamics.
\emph{Rev.\ Mod.\ Phys.\ } \textbf{1949}, {\textit {21}}, 392--399.

\bibitem{74}
Czachor, M.; Wrzask, K.
Automatic regularization by quantization in reducible representations
of CCR: Point-form quantum optics with classical sources in the Milne
universe.
\emph{Int.\ J.\ Theor.\ Phys.\ } \textbf{2009}, {\textit {48}}, 2511--2549.

\bibitem{75}
Wald, R.M.
\emph{General Relativity}; The university of Chicago Press: Chicago, IL, USA, 
1984.

\bibitem{76}
Rindler, W.
Kruskal Space and the Uniformly Accelerated Frame.
\emph{Am.\ J.\ Phys.\ } \textbf{1966}, {\textit {34}}, 1174--1178.

\bibitem{77}
Koks, D.
\emph{Explorations in Mathematical Physics: The Concepts Behind an Elegant
Language}; Springer: New York, NY, USA, 2006.

\bibitem{78}
Culetu, H.
Kinematic parameters in the spherical Rindler frame spacetime. \textbf{2008},
{arXiv:0804.3754.}

\bibitem{79}
St. Augustine of Hippo
\emph{The Confessions of St. Augustine}; Translated by Sheed, F. J.;
Sheed \& Ward: New York, NY, USA, 1943.

\bibitem{80}
Vilenkin, A.
Quantum Origin of the Universe.
\emph{Nucl.\ Phys.\ B} \textbf{1985}, {\textit {252}}, 141--151.

\bibitem{81}
Vilenkin, A.
\emph{Many Worlds in One: The Search for Other Universes};
Hill \& Wang: New York, NY, USA, 2006.

\bibitem{82}
St. Augustine of Hippo \emph{City of God};
Penguin Books: London, UK, 1984. (Translated by {Henry Bettenson})

\bibitem{83}
Hawking, S.W.
The edge of spacetime. In  \emph{The New
Physics}; Davies, P., Ed.; Cambridge University Press: Cambridge, UK, 1989; 
p. 69.

\bibitem{84}
Gr\"{u}nbaum, A.
Creation As a Pseudo-Explanation in Current Physical Cosmology.
\emph{Erkenntnis} \textbf{1991}, {\textit {35}}, 233--254.

\bibitem{85}
Hawking, S.W.
Quantum Cosmology. In
\emph{Three Hundred Years of Gravitation}; Hawking, S.W.; Israel, W.; Eds.; 
Cambridge University Press: Cambridge, UK, 1987; pp. 631--651.

\bibitem{86}
Fl\"{u}gge, S.
\emph{Practical Quantum Mechanics,Vol. 1};
Springer: Berlin, Germany, 1994; pp. 196--197.

\bibitem{87}
Luke, Y.L.
\emph{Integrals of Bessel Functions}; McGraw-Hili Book Co.: New York, NY, USA, 
1962.

\bibitem{88}
Birrell, N.D.; Davies, P.C.W.
\emph{Quantum Fields in Curved Space}; Cambridge University Press:
Cambridge, UK, 1982.

\bibitem{89}
Ford, L.H.
Quantum field theory in curved space-time.
In
\emph{Particles and Fields.} Proceedings of the 9th Jorge Andre Swieca Summer
School, Campos do Jordao, Brazil, 1997; Barata, J.C.A., Malbouisson, A.P.C., 
Novaes, S.F.; Eds.; World Scientific: Singapore, 1998;
pp. 345--388.

\bibitem{90}
Schutz, B.F.
\emph{A First Course in General Relativity}; Cambridge University Press:
Cambridge, UK, 2009; p. 153.

\bibitem{91}
Padmanabhan, T.
Physical interpretation of quantum field theory in noninertial
coordinate systems.
\emph{Phys.\ Rev.\ Lett.\ } \textbf{1990}, {\textit {64}}, 2471--2474.

\bibitem{92}
Tolley, A.J.; Turok, N.
Quantum fields in a big crunch / big bang space-time.
\emph{Phys.\ Rev.\ D} \textbf{2002}, {\textit {66}}, 106005.

\bibitem{93}
Parker, L.
Quantized fields and particle creation in expanding universes.
\emph{Phys.\ Rev.\ } \textbf{1969}, {\textit {183}}, 1057--1068.

\bibitem{94}
Chitre, D.M.; Hartle, J.B.
Path Integral Quantization and Cosmological Particle Production:
An Example.
\emph{Phys.\ Rev.\ D} \textbf{1977}, {\textit {16}}, 251--260.

\bibitem{95}
Watson, G.N.
\emph{A Treatise on the Theory of Bessel Functions}; Cambridge University
Press: Cambridge, UK, 1922; p. 180.

\bibitem{96}
Nikishov, A.I.; Ritus, V.I.
Rindler solutions and their physical interpretation.
\emph{J.\ Exp.\ Theor.\ Phys.\ } \textbf{1998}, {\textit {87}}, 421--425.

\bibitem{97}
Nikishov, A.I.; Ritus, V.I.
Processes induced by a charged particle in an electric field, and the
Unruh heat-bath concept.
\emph{Sov.\ Phys.\ JETP } \textbf{1988}, {\textit {68}}, 1313--1321.

\bibitem{98}
Bili\'{c}, N.; Toli\'{c}, D.
FRW universe in the laboratory.
\emph{Phys.\ Rev.\ D} \textbf{2013}, {\textit {88}}, 105002.

\bibitem{99}
Custodio, P.S.; Horvath, J.E.
The Evolution of primordial black hole masses in the radiation
dominated era.
\emph{Gen.\ Rel.\ Grav.\ } \textbf{2002}, {\textit {34}}, 1895--1907.

\bibitem{100}
Arp, H.
Ambartsumian's greatest insight---the origin of galaxies. In
\emph{Active Galactic Nuclei and Related Phenomena};
Proceedings of the International Astronomical Union Syposium 194, held 17-21 
August, 1998, in Yerevan, Armenia; Terzian, Y.; Weedman, D.; Khachikian, E.;
Eds.;  Astronomical Society of the Pacific: San Francisco, USA, 1999;
pp. 473--477.

\bibitem{101}
Vishwakarma, R.G.
A curious explanation of some cosmological phenomena.
\emph{Phys.\ Scripta} \textbf{2013}, {\textit 5}, 055901.

\bibitem{101A}
Chodorowski, M.J.
Cosmology under Milne's shadow. \emph{Publ.\ Astron.\ Soc.\ Austral.\ } 
\textbf{2005}, {\textit {22}}, 287--291.


\bibitem{102}
Bondi, H.
\emph{Cosmology}; Cambridge University Press: Cambridge, UK, 1960.

\bibitem{103}
Gale, G.
Cosmology: Methodological Debates in the 1930s and 1940s. In
 \emph{The Stanford Encyclopedia of Philosophy (Spring
2014 Edition)}; Zalta, E.N., Ed.
Available online: 
http://plato.stanford.edu/archives/spr2014/entries/cosmology-30s/ 
(accessed on 21.09.2015).

\bibitem{104}
Lepeltier, T.
Edward Milne's influence on modern cosmology.
\emph{Ann. Sci.} \textbf{2006}, {\textit{ 63}}, 471--481.

\bibitem{105}
Urani, J.; Gale, G.
E.A. Milne and the Origins of Modern Cosmology: An Ubiquitous Presence.
In
\emph{The Attraction of Gravitation: New Studies in the History of General
Relativity}; Earman, J.; Janssen, M.; Norton, J.D.; Eds.; Birkhaeuser: Boston, 
MA, USA, 1994; pp. 390--419.

\bibitem{106}
Walker, A.G.
The Principle of Least Action in Milne's Kinematical Relativity.
\emph{Proc.\ Roy.\ Soc.\ Lond.\ A} \textbf{1934}, {\textit {147}}, 478--490.

\bibitem{107}
Walker, A.G.
On the formal comparison of Milne's kinematical system with the systems
of general relativity.
\emph{Mon.\ Not.\ Roy.\ Astron.\ Soc.\ } \textbf{1935}, {\textit {95}}, 
263--269.

\bibitem{108}
Ingarden, R.S.
On physical applications of Finsler geometry.
\emph{Contemp.\ Math.\ } \textbf{1996}, {\textit {196}}, 213--223.

\bibitem{109}
Antonelli, P.L.; Ingarden, R.S.; Matsumoto, M.
\emph{The Theory of Sprays and Finsler Spaces with Applications in Physics
and Biology}; Kluwer Academic Publishers: Dordrecht, The \mbox{Netherlands, 
1993.}

\bibitem{110}
Infeld, L.; Schild, A.
A New Approach to Kinematic Cosmology.
\emph{Phys.\ Rev.\ } \textbf{1945}, {\textit {68}}, 250--272.

\bibitem{111}
Gr\o n, O.; Johannesen, S.
FRW Universe Models in Conformally Flat Spacetime Coordinates. II:
Universe models with negative and vanishing spatial curvature.
\emph{Eur.\ Phys.\ J.\ Plus}\textbf{ 2011}, \mbox{{\textit {126}}, 29.}


\bibitem{112}
Eisenhart, L.P.
\emph{Riemannian Geometry}; Princeton University Press: Princeton, NJ, USA, 
1949; \mbox{p. 188.}

\bibitem{113}
Robertson, H.P.
Relativistic Cosmology.
\emph{Rev.\ Mod.\ Phys.\ } \textbf{1933}, {\textit 5}, 62--90.

\bibitem{114}
Gogberashvili, M.
Our world as an expanding shell.
\emph{Europhys.\ Lett.\ } \textbf{2000}, {\textit {49}}, 396--399.

\bibitem{115}
Randall, L.; Sundrum, R.
An Alternative to compactification.
\emph{Phys.\ Rev.\ Lett.\ } \textbf{1999}, {\textit {83}}, 4690--4693.

\bibitem{116}
Arkani-Hamed, N.; Dimopoulos, S.; Dvali, G.R.
The Hierarchy problem and new dimensions at a millimeter.
\emph{Phys.\ Lett.\ B} \textbf{1998}, {\textit {429}}, 263--272.

\bibitem{117}
Penrose, R.
The basic ideas of conformal cyclic cosmology.
\emph{AIP Conf.\ Proc.\ } \textbf{2012}, {\textit {1446}}, 233--243.

\bibitem{118}
Khoury, J.; Ovrut, B.A.; Steinhardt, P.J.; Turok, N.
The Ekpyrotic universe: Colliding branes and the origin of the hot big
bang.
\emph{Phys.\ Rev.\ D} \textbf{2001}, {\textit {64}}, 123522.

\bibitem{119}
Steinhardt, P.J.; Turok, N.
Cosmic evolution in a cyclic universe.
\emph{Phys.\ Rev.\ D} \textbf{2002}, {\textit {65}}, 126003.

\bibitem{120}
Carr, B.J.; Coley, A.A.
Persistence of black holes through a cosmological bounce.
\emph{Int.\ J.\ Mod.\ Phys.\ D} \textbf{2011}, {\textit {20}}, 2733--2738.

\bibitem{121}
Khoury, J.; Ovrut, B.A.; Seiberg, N.; Steinhardt, P.J.; Turok, N.
From big crunch to big bang.
\emph{Phys.\ Rev.\ D} \textbf{2002}, {\textit {65}}, 086007.


\end{thebibliography}
\end{document}